\documentclass[12pt]{article}
 \pdfoutput=1
\usepackage{jcappub}
\usepackage[top=1in, bottom=1in, left=1in, right=1in]{geometry}
\usepackage[font=footnotesize]{caption}
\usepackage{array}
\usepackage{enumerate}
\usepackage{float}
\usepackage{subfig}
\usepackage{multicol}
\usepackage{multirow}
\usepackage{epsfig}
\usepackage{graphicx}
\usepackage{pict2e}
\usepackage{color}
\usepackage{amsmath,amsfonts,amssymb}
\usepackage{bbm}
\usepackage[utf8]{inputenc}
\usepackage[english]{babel}
\usepackage{xspace}

\newcommand{\be}{\begin{equation}}
\newcommand{\ee}{\end{equation}}
\newcommand{\bea}{\begin{eqnarray}}
\newcommand{\eea}{\end{eqnarray}}

\newcommand{\vp}{\varphi}
\newcommand{\dvp}{\delta\varphi_1}
\newcommand{\dvpI}{\delta\varphi_{1I}}
\newcommand{\dvpK}{\delta\varphi_{1K}}
\newcommand{\Hcal}{\mathcal H}
\newcommand{\ddvpI}{\delta \varphi_{2I}}
\newcommand{\ddvpK}{\delta \varphi_{2K}}
\newcommand{\ddvp}{\delta \varphi_{2}}

\def\Xkdvk{{\sum_K X_K \delta\vp_{1K}}}
\title{Sourced Scalar Fluctuations in Bouncing Cosmology}

\begin{abstract}
{We calculate the scalar power spectrum generated by sourced fluctuations due to coupling between the scalar field, which holds most of the energy density of the universe, and a gauge field for a general FLRW metric. For this purpose we calculate the curvature perturbation to second order in the presence of gauge fields, and show that the gauge fields behave like an additional potential term. We then apply the analysis to the case of slow-contraction. Due to the interaction between the scalar field and gauge fields additional 'sourced' tensor and scalar spectra are generated. The resulting spectra are chiral, slightly blue and arbitrarily close to scale invariance. The only difference between the tensor and scalar spectra is the coupling constant with an ${\mathcal O}(1)$ numerical coefficient, and some momentum space polarization vectors. As a result the tilt of the spectra are the same. For the nearly scale invariant case, the momentum integration gives the same leading contribution. Hence, $r\simeq 1/9$ where the deviation from this value is controlled by the deviation from scale invariance, and is not in agreement with CMB observations. 
Deviating considerably from near scale invariance, and considering a bluer tilt with $n_T>0.12$, the model cannot account for CMB observations, but can be detected by LIGO and/or LISA in the future.}
\end{abstract}
\author{Ido Ben-Dayan$^{1}$, Judy Kupferman$^{2}$}
\affiliation{$^1$ Physics Department, Ariel University, Ariel 40700, Israel\\
$^2$Center for Quantum Science and Technology,
Ben-Gurion University of the Negev, P.O. Box 653, Be'er-Sheva 8410500, Israel
}
\emailAdd{ido.bendayan@gmail.com,judithku@post.bgu.ac.il}
\begin{document}

\maketitle
\section{Introduction}
Current CMB observations of large scales in the Universe have measured a slightly red tilted spectrum of scalar fluctuations and have placed an upper bound on the tensor to scalar ratio of $r\leq 0.06$ for the $\Lambda$CDM concordance model \cite{Akrami:2018odb}. The inflationary paradigm generically predicts a nearly scale invariant spectrum, where the value of $r$ is a model dependent statement. Therefore, as a paradigm, Inflation provides an excellent match to the observed CMB data. For generic slow-roll models, $r$ is tied to the energy scale of inflation, giving us an invaluable handle on Physics near the GUT scale. The progressing bounds on $r$ over the past twenty years have ruled out simple models, including a linear potential \cite{Akrami:2018odb}. In addition to the continuing but slow shrinkage in model space, Inflation does not resolve the Big Bang singularity \cite{Borde:2001nh}. It is therefore useful to consider alternatives that resolve the Big Bang singularity such as bouncing models.  

The recent observation of gravitational waves (GW) with LIGO \cite{Abbott:2016blz,TheLIGOScientific:2016dpb} opens up a new possibility of measuring primordial GW. Such laser interferometers probe much smaller scales than CMB observations. However, both LIGO and LISA sensitivity bands are order of magnitudes weaker than CMB sensitivity.
The inflationary slow-roll prediction, of slightly red tensor spectrum, means that the Laser Interferometer (LI) observations will come out empty handed if they ever surmount the experimental difficulties of cleaning the noise from the signal. Thus, if Inflation is the actual paradigm realized in Nature, and LI measure a primordial GW signal, it can only be due to some New Physics beyond $\Lambda$CDM and Inflation. For instance, on the potential of LISA discoveries, see a recent review \cite{Bartolo:2016ami}.

A celebrated example that deviates from the standard slow-roll predictions on various scales and observables are models with sourced fluctuations, along the lines of \cite{Sorbo:2011rz,Barnaby:2010vf, Barnaby:2011vw,Barnaby:2012tk, Caprini:2014mja,jiro1,jiro2}. In these models, the coupling between the inflaton and a gauge field generates sourced fluctuations that account for rich phenomena. For example, the sourced fluctuations generate additional spectra on top of the adiabatic one and disentangle the link between the energy scale of inflation and $r$.  
Nevertheless, the potential discovery of LI points again at considering alternatives for Inflation, and not just modifying the inflaton or matter lagrangian. 

With these motivations in mind, we consider bouncing models that provide a different approach. There is no Big Bang singularity, and in return, one usually invokes a temporary null energy condition violation. The outcomes are rather delicate models, prone to various instabilities \cite{Battefeld:2014uga, Cai:2007qw, Cai:2008qw, Lehners:2010fy, Easson:2011zy, Cai:2012va, Cai:2014bea, Cai:2016thi}. Two characteristic scenarios are slow contraction like the Pre Big Bang \cite{Gasperini:2002bn} or ekpyrotic model \cite{Lehners:2010fy}, and a matter bounce, where the contracting phase is dominated by matter. 
In the ekpyrotic case, the Universe starts with a slow contraction, followed by kinetic dominated contraction, a bounce, kinetic dominated expansion and the reheating to the Hot Big Bang scenario  \cite{Lehners:2010fy}. The main motivation for considering slow contraction, rather than a matter bounce, is due to the anisotropic instability \cite{Battefeld:2014uga}. Excluding the matter bounce, single field bouncing models predict a blue scalar and tensor spectra.
To conform with CMB observations of a slightly red tilted scalar spectrum, an entropic/curvaton mechanism is invoked, usually by introducing an additional scalar field. The tensor spectrum is left unchanged \cite{Creminelli:2014wna,Geshnizjani:2014bya}, i.e. a very blue spectrum, with $n_T\sim 2-3$. Hence, LI observations of the next twenty years might observe such a signal \cite{Veneziano:1991ek, Gasperini:1992em, Gasperini:2002bn, Gasperini:1992dp, Brustein:1995ah, Brustein:1996ut, Starobinsky:1979ty,Boyle:2003km, Gasperini:2016gre},  while the signal is orders of magnitude below any devised CMB experiment. 

We are interested in considering the fruitful sourced fluctuations idea with the contracting background. Deviating from the scalar field(s) framework of bouncing models, and considering sourced fluctuations, the above stated predictions change considerably  \cite{Chowdhury:2016aet, Chowdhury:2015cma, Chowdhury:2018blx,Gasperini:2017fqw,Ben-Dayan:2016iks}.
In \cite{Ben-Dayan:2016iks} it was demonstrated for the first time that even in a bouncing model, a nearly scale invariant tensor spectrum can be generated due to sourced fluctuations. The tensor spectrum is \textit{slightly blue} $0<n_T\leq 0.3$, and \textit{chiral}. These features make such a spectrum rather unique, and in principle an easy target for detection. The bounds on $n_T$ come from backreaction constraints and assuming a level of tensor to scalar ratio $r>10^{-4}$. Such a sourced spectrum could be observed by CMB observations and by LI observations if $n_T\sim 0.3$ \cite{Boyle:2007zx}.

However, \cite{Ben-Dayan:2016iks} assumed that the scalar spectrum is somehow being generated in the right amount, such that it fits CMB observations. As such, \cite{Ben-Dayan:2016iks} was a proof of concept, rather than a competitive model.\footnote{An attempt calculating the scalar and tensor spectrum for a related model $\gamma=0$ has been carried out in \cite{Ito:2016fqp} yielding $r\simeq 7$. Our calculation disagrees with their results. We will show that the model yields $r\sim 1/9$, as with the $\gamma\neq 0$ case.}  
In this work, we close the gap by calculating the scalar spectrum generated by sourced fluctuations in the model discussed by \cite{Ben-Dayan:2016iks} and hence the value of $r$.

Calculation of the scalar spectrum in bouncing models has been plagued by gauge artifacts \cite{Creminelli:2004jg}. To avoid that, we carried out a full second order derivation of the metric and field equations, as means of extracting the correct source term.
 We generalize previous derivations to any flat FLRW geometry, multiple scalar fields and in the presence of gauge fields. We show that the gauge fields appear at the second order equations in a manner similar to a potential term. This should be true in general for any field that gives a negligible contribution to the background and appears quadratically or with higher power in the action. As a result, the sole difference between the tensor and scalar fluctuations is an $\mathcal O(1)$ factor in the coupling constants and some phase space factors.
 At the limit of scale invariance, we find that $r\sim 1/9$, that is ruled out by observations. The simplest model of inflation with sourced fluctuations \cite{Barnaby:2010vf} has also been ruled out by non-gaussianity bounds. Therefore, it seems that while sourced fluctuations are a good alternative for generating primordial spectra, an actual viable realization, whether inflation or a bounce, is rather non-trivial. 
 
 Finally, if we deviate from scale invariance and the idea that the model is responsible for CMB observations, we can consider a bluer spectrum with $n_T>0$. In such case, we find that for some range of parameters the model generates a GW spectrum that is observable by LISA and/or LIGO.\footnote{We thank the anonymous referee for posing the question of whether the model is detectable by LI observations.}

The paper is organized as follows. We begin by describing the setup of the model, and mention some previous results, such as the solutions for the gauge fields and the backreaction bounds. In section $3$ we give the major steps leading to the equation for the gauge invariant curvature perturbation. The full derivation is given in Appendix A. In section $4$ we perform the calculation of the spectra and tensor to scalar ratio. An example of the momentum integral calculation is relegated to Appendix B. In section $5$ we analyze the case of the predictions of the model for present day GW searches assuming that CMB measurements are explained by some other means.  We then conclude. 

\section{Setup and Previous Analysis}
The scenario we are interested in is a scalar field $\varphi$ coupled to some $U(1)$ gauge field, $A_{\mu}$ with the action: 
\begin{align}
S & =\int d^{4}x\sqrt{-g}\left[\frac{M_{pl}^{2}}{2}R-\frac{1}{2}\left(\partial\varphi\right)^{2}-U\left(\varphi\right)-I^{2}\left(\tau\right)\left\{ \frac{1}{4}F^{\mu\nu}F_{\mu\nu}-\frac{\gamma}{4}\tilde{F}^{\mu\nu}F_{\mu\nu}\right\} \right]\label{eq:action},
\end{align}
where $I\left(\tau\right) =\left(-\tau\right)^{-n}\equiv a_{2}^{n}e^{-n\varphi/a_{1}},\, a_1=\sqrt{2p}M_{pl}/(1-p)\sim \sqrt{2p}M_{pl}$ 
 \footnote{We correct the way $a_1$ was expressed in \cite{Ben-Dayan:2016iks}. In  \cite{Ben-Dayan:2016iks}, this error did not interfere with the calculation and the final result, while here $a_1$ does go into the calculation of the scalar spectrum and as a result to the derivation of $r$.} and $a_2$ can be read off by equating $\varphi\equiv a_1\ln(-a_2\tau)$ with equation \eqref{eq:phi one term} below. Another possible function, such as $I(\tau)\sim \varphi^n$, would add logarithmic corrections to the behavior discussed below, but the qualitative behavior will remain unchanged. If $\gamma$ is indeed a parameter then parity is explicitly broken. However it can easily be a vev of some pseudoscalar (See for example \cite{Caprini:2014mja}.) 
\subsection{Background Solution}
Writing the flat FLRW metric in cosmic  $(t)$ and conformal time $(\tau)$:
\be
ds^2=-dt^2+a^2(t) d\vec {x}^2=a^2(\tau)\left[-d\tau^2+d\vec {x}^2\,\right],
\ee
a dot denotes differentiation with respect to cosmic time and prime denotes a differentiation in conformal time. In the absence of gauge fields, an exact scaling solution of the equations of motion is given by:
\begin{align}
U=-U_{1}e^{-\sqrt{2/p}\varphi},\:a\left(t\right)\sim\left(-t\right)^{p},\:-\infty\leq t\leq0,
\end{align}
$\varphi_{1}$ is written in Planck units and $p>0$ is dimensionless.
Power law inflation corresponds to $p\gg1,$ $U_{1}<0$, the matter
bounce to $p=2/3,\,U_{1}>0$, and  $p\ll1,$ $U_{1}>0$ to ekpyrosis. The generalization to several fields
is:
\begin{align}
a=\left(-t\right)^{p},\,\,\:p\ll1,\,\,\:p=\sum_{i}\frac{2}{c_{i}^{2}},\,\,\:H=\frac{p}{t}
\end{align}
\begin{align}
 \varphi_{i}=\frac{2}{c_{i}}ln\left(-\sqrt{\frac{U_{i}}{2/c_{i}^{2}\left(1-3p\right)M_{pl}^{2}}}t\right),\,\,\:\dot{\varphi_{i}}=\frac{2}{c_{i}t},\,\,\:\ddot{\varphi}_{i}=-\frac{2}{c_{i}t^{2}},\label{eq:phi one term}
\end{align}
\begin{align}
U=-\sum_{i}U_{i}e^{-c_{i}\varphi_{i}}=-\frac{p\left(1-3p\right)}{t^{2}}.
\end{align}
We will use conformal time, $-\infty<-\tau\leq0$ with $-\tau=\left(-t\right)^{1-p}/\left(1-p\right)$.
Then the scale factor is
\begin{align}
a\left(t\right)\sim\left(-t\right)^{p}=\left(-\left(1-p\right)\tau\right)^{p/\left(1-p\right)},\:\mathcal{H}=\frac{p}{\left(1-p\right)\left(-\tau\right)}
\end{align}
\begin{align}
\varphi_{i}'=\frac{2}{c_{i}\left(1-p\right)\tau},\:\varphi_{i}^{''}=-\frac{2}{c_{i}\left(1-p\right)\tau^{2}},\:U=-\frac{p\left(1-3p\right)}{\left(-\left(1-p\right)\tau\right)^{2/\left(1-p\right)}}.
\end{align}
We normalize the scale factor so that it is unity at the end of ekpyrosis,
$a\left(\tau\right)=\left(-\tau/\tau_{end}\right)^{b}$, and $b\equiv p/\left(1-p\right).$
In ekpyrosis, $p\ll1$, so $b\simeq p.$

\subsection{Scalar and Tensor Perturbations}
Denoting $\hat X=\hat \zeta, \hat h$ as the curvature and tensor perturbation respectively, and defining $\hat Q_k=a\hat X$, the equation of motion for the perturbation is:
 \begin{equation}
\left[ \partial_\tau^2 + \left( k^2 - \frac{f''}{f} \right) \right]  Q_\lambda \left( \tau ,\, \vec{k} \right) = J_\lambda \left( \tau ,\, \vec{k} \right)
\label{hc-eq-formal}
\end{equation} 
where $J_{\lambda}$ is a source term due to the presence and interaction of gauge fields and $\lambda$ are the helicity eigenstates of +,-.  In the case of vacuum fluctuations $J\equiv 0$. $f=a(\tau)$ for the tensor perturbation and $f=-\frac{H}{\dot{\varphi}}a(\tau)$ for the curvature perturbation. In our specific model $-H/\dot{\varphi}=\sqrt{p/2}$, hence $f''/f=a''/a$ in both cases.
Decomposing into vacuum and sourced fluctuations:
\begin{align}
Q_k \left( \tau \right)&=Q^v_{\vec k}(\tau)+Q^s_{ \vec k}(\tau)\\
Q_k^v\left( \tau \right)&=b(\vec k)f_k(\tau)+b^{\dagger}(-\vec k)f^*_k(\tau),\quad \left[b(\vec k),b^{\dagger}(\vec k')\right]=\delta^{(3)}(\vec k-\vec k').
\end{align}
The power spectrum is
related to the correlator via
\begin{align} \label{eq:correlator}
<\hat X_{k}\hat X_{k'}>=\frac{2\pi^{2}}{k^3}\delta(\vec{k}+\vec{k}')({\cal P}^v_X(k)+{\cal P}^s_X(k))
\end{align}
where $\hat X_k$ denotes the curvature (scalar) and tensor perturbations. 
The power spectrum for scalar perturbation $\hat \zeta$ is
\begin{align}
{\cal P}_{S}(k)\simeq A_{S}\left(\frac{k}{k_{0}}\right)^{n_{S}-1}
\end{align}
where $n_S\simeq 0.97$ is known as the spectral tilt, the amplitude is measured to be $A_{S}\simeq 2.1\times 10^{-9}$ \cite{Akrami:2018odb}. For gravitational waves, $\hat h$, the power spectrum is
\begin{align}
{\cal P}_{T}(k)\simeq A_{T}\left(\frac{k}{k_{0}}\right)^{n_{T}}
\end{align}
and $r\equiv A_{T}/A_{S}$ with current constraints $r<0.06$ \cite{Akrami:2018odb}. 
In ekpyrosis, with $p \ll1$ the spectra are very blue $\mathcal{P}\sim k^2$ giving negligible contribution on CMB scales. Hence, bouncing models need an additional source to generate the measured scalar spectrum. The inclusion of gauge fields provides a natural candidate for such a source. The tensor spectrum due to sourced fluctuations from \eqref{eq:action} was calculated in \cite{Ben-Dayan:2016iks}, allowing $0<n_T\lesssim 0.3$ in accord with current data:
 \be
 {\cal P}_T^s\simeq  \frac{11.1}{256\pi^6 n_T}\frac{e^{4\pi \xi}}{b^{4}\xi^{6}}\left(\frac{H_{end}}{M_{pl}}\right)^4\left(\frac{k}{H_{end}}\right)^{n_T}
 \ee
where $\xi\equiv -\gamma n$, with $n_T=4(2+n)$ for $n<-1/2$ and $n_T=4(1-n)$ for $n>-1/2$, and $H_{end}$ is the Hubble parameter at the end of the slow contraction. Note that $n=-2,1$ imply a scale invariant spectrum, and the deviation from these values control the deviation from a scale invariant spectrum.
We will calculate the scalar spectrum resulting from the action with gauge fields,
and the resulting r can be compared to observations.

\subsection{Inclusion of gauge fields}
In order to deal with the source term that will appear in the equation of motion, we redefine the gauge field,  $\tilde{A}=IA$.
In the Coulomb gauge, $A_{0}=\partial^{i}A_{i}=0$, for the canonically
normalized field, the lagrangian is \cite{Caprini:2014mja}:

\begin{equation}
\mathcal{L}=\frac{1}{2}\tilde{A'}_{i}^{2}-\frac{1}{2}\frac{I''}{I}\tilde{A}_{i}^{2}-\gamma\frac{I'}{I}\epsilon_{ijk}\tilde{A}_{i}\partial_{j}\tilde{A}_{k}.
\end{equation}
Note that this term is invariant under 
\begin{equation}
n\rightarrow-1-n,\:\gamma\rightarrow-\gamma\frac{n}{1+n}.\label{eq:transformation}
\end{equation}
Notice that $\xi$ is also invariant under these transformation, as follows:
\begin{equation}
\xi\equiv-n\gamma\rightarrow\xi.
\end{equation}
$\xi$ parametrizes the enhancement of the gauge field fluctuations and hence the scalar and tensor sourced fluctuations spectra.
 The gauge field operator is decomposed according to:
 \begin{equation}
  \vec{A}(\tau,{\vec x}) = \sum_{\lambda=\pm} \int \frac{d^3k}{(2\pi)^{3/2}} \left[ \vec{\epsilon}_\lambda({\vec k}) a_{\lambda}({\vec k}) A_\lambda(\tau,{\vec k}) e^{i {\vec k}\cdot {\vec x}} + \mathrm{h.c.}   \right]
\label{decomposition}
\end{equation}
with the standard commutation relations:
\begin{equation}
  \left[a_{\lambda}({\vec k}), a_{\lambda'}^{\dagger}({\vec k'})\right] = \delta_{\lambda\lambda'}\delta^{(3)}({\vec k}-{\vec k'}).
\label{ladder}
\end{equation}
The polarization vectors $\vec{\epsilon}_\lambda$ fulfill  $\vec{k}\cdot \vec{\epsilon}_{\pm} \left( \vec{k} \right) = 0$, 
$\vec{k} \times \vec{\epsilon}_{\pm} \left( \vec{k} \right) = \mp i k \vec{\epsilon}_{\pm} \left( \vec{k} \right)$,
$\vec{\epsilon}_\pm \left( \vec{-k} \right) = \vec{\epsilon}_\pm \left( \vec{k} \right)^*$, and are normalized according to $\vec{\epsilon}_\lambda \left( \vec{k} \right)^* 
\cdot \vec{\epsilon}_{\lambda'} \left( \vec{k} \right) = \delta_{\lambda \lambda'}$.  
The annihilation and creation operators of the gauge field commute with the operators of the tensor and scalar fluctuations:
\be
\label{abcommute}
 \left[b({\vec k}), a_{\lambda'}^{\dagger}({\vec k'})\right] =\left[b({\vec k}), a_{\lambda'}({\vec k'})\right]=0
\ee 
which is the reason why there are no cross terms in \eqref{eq:correlator}.
The mode functions $\tilde{A}_{\lambda}$ are the solution of:
\begin{equation}
\tilde{A}{}_{\lambda}^{''}+\left(k^{2}+2\lambda\xi\frac{k}{\tau}-\frac{n\left(n+1\right)}{\tau^{2}}\right)\tilde{A}_{\lambda}=0.\label{eq:EOM for A}
\end{equation}
With the boundary conditions of Bunch-Davies vacuum as $\tau\rightarrow -\infty$ the solutions are Coulomb wave functions:
\be
\tilde A=\frac{1}{\sqrt{2k}}\left(G_{-n-1}(\xi,-k \tau)+i F_{-n-1}(\xi,-k\tau)\right).
\ee
For our purposes, the important region is outside the horizon as $-k\tau \ll 1/\xi \ll 1$ and for $\lambda=+$ that is enhanced by a factor of $e^{\pi \xi}$: 
\bea \label{eq:modeA}
\tilde {A}_{\lambda} (\tau, \vec k)&\simeq& \sqrt{-\frac{\tau}{2\pi}}e^{\xi\pi}\Gamma(|2n+1|)|2\xi k \tau|^{-|n+1/2|}
\eea
and will be used in section 4 to express the source term in the equation of motion, and obtain the curvature perturbation.

Considering the additional gauge fields, one needs to verify that they do not dominate over the scalar field in charge of the slow contraction. The analysis was carried out in \cite{Ben-Dayan:2016iks}. It limits the parameter $n$ to be between $-2<n<1$, otherwise the energy density of the gauge fields diverges and the slow contraction analysis is not valid. Furthermore, it constrains the Hubble parameter $H$ during the slow contraction to be:
\be
\label{eq:backreaction}
   H/M_{pl}\ll \sqrt{3/D_{1,2}(n)} \,p^2\xi^{3/2}e^{-\pi \xi}, \quad D_2(n)\equiv\frac{1}{4\pi^2} \frac{  (n+1)^2 \Gamma (-2 n-1)^2}{2^{1-2 n}\pi  (n+2)}.
 \ee 
where $D_2(n)$ refers to $-1/2>n>-2$ and $D_1(n)$ is obtained by the substitution of $n\rightarrow-1-n$ and is relevant for $1>n>-1/2$. The result obtained in \cite{Ben-Dayan:2016iks} of $0<n_T\lesssim 0.3$ is in accordance with this backreaction bound.

\section{Second Order Klein-Gordon Equation}
Let us derive the second order differential equation for the curvature perturbation. In this section we report mostly the single field result as we shall only consider a single field in our analysis. Nevertheless, we provide the full derivation of the multi-field case in Appendix A. The derivation is valid for any FLRW metric, without any fast-roll or slow-roll approximations. Working in the flat gauge, we closely follow the derivation by Malik in \cite{Malik:2006ir}, with the addition of gauge fields at second order.
 We use natural units where $8 \pi G=M_{pl}^{-2}=1$. The action becomes
\be
\label{action}
\mathcal{S}=\int d^4x\sqrt{-g}\left[\frac{R}{2}-\frac{1}{2}(\partial \varphi)^2-U(\varphi)-\frac{I^2(\varphi)}{4}(F^2-\gamma F \tilde F)\right].
\ee

The Klein-Gordon (KG) equation is given by:
\be 
\frac{1}{\sqrt{-g}}\partial_{\mu}\left(\sqrt{-g}g^{\mu \nu}\partial_{\nu}\varphi\right)+\frac{dU}{d\varphi}=-\frac{1}{4}\frac{dI^2}{d\varphi}(F^2-\gamma F \tilde F).
\ee

As long as \eqref{eq:backreaction} is fulfilled, we can neglect the gauge field contribution to the background. In the KG equation the first order contribution $\delta A_{\mu}$ appears only quadratically, so it is a second order contribution. Hence, in deriving the KG equation in closed form, there will be no changes in the zeroth and first order equations.  For bookkeeping the gauge field term behaves as a second order potential term, and therefore will only change the diagonal terms of the Einstein field equations (EFE) at second order.
\bea 
U(\varphi)+\frac{I^2(\varphi)}{4}(F^2-\gamma F \tilde F)=U_0+\delta U_1+\frac{1}{2}\delta U_2+\frac{I^2(\varphi_0)}{4}(F^2-\gamma F \tilde F)\cr
=U_0+U_{\varphi}\delta \varphi_1+\frac{1}{2}(U_{\varphi \varphi}\delta \varphi_1^2+U_{\varphi}\delta \varphi_2)+\frac{I^2(\varphi_0)}{4}(F^2-\gamma F \tilde F)
\eea
where $U_0\equiv U(\varphi_0)$ and  
$U_{\varphi}=\frac{\partial U}{\partial \varphi},U_{\varphi \varphi}=\frac{\partial^2 U}{\partial \varphi^2}$ etc.

We would like to express the evolution equation of the fluctuations using just the scalar field fluctuations $\delta \varphi$ and gauge field fluctuations $\delta A_{\mu}$. Then using the gauge invariant quantity for the curvature perturbation $\zeta=-\frac{H}{\dot{\varphi_0}}\delta \varphi=-\frac{\Hcal}{\varphi_0'}\delta \varphi$, we will calculate the scalar power spectrum of the sourced fluctuations.
%
\subsection{Zeroth and First Order}\label{0th}
The Klein-Gordon equation at zeroth order is:
\be
\varphi_{0}''+2\mathcal{H} \varphi_{0}'+a^2U_{\varphi}=0
\label{flatKG0real}
\ee
Using the EFE, we get an equation for the first order field fluctuation which has no metric fluctuations in it, but just field fluctuations and background quantities:
%
\be
\label{flatKG1real}
\delta \vp_{1}''+2\Hcal\delta \vp_{1}'-\nabla^2\delta \vp_{1}
+a^2\left\{
U_{\vp \vp}
+\frac{1}{\Hcal}\left(
2\vp_{0}'U_{\vp}
+\frac{\vp_{0}'^2}{\Hcal}U_0
\right)
\right\}\delta \vp_{1}=0\,.
\ee

These first order fluctuations generate the well known vacuum fluctuations of a slowly contracting Universe with a blue spectrum $P_S\sim k^2$.
As an indication that our analysis is correct, we find that the entire term in the curled brackets vanishes. 
This confirms the known result that in a slowly contracting Universe, one cannot neglect the metric perturbations as they exactly cancel the contribution from the potential, and the behavior of the curvature perturbation, $\zeta$ is that of a massless free field \cite{Creminelli:2004jg}.
The gauge fields  from \eqref{action} will appear at 2nd order in the KG equation as a source term, hence their name 'sourced fluctuations'.

\subsection{Second Order}
The 2nd order KG equation in the flat gauge for the multi-field case in the presence of gauge fields is given by:
\bea
\label{flatKG2real}
\ddvpI''&+&2\Hcal\ddvpI'-\nabla^2\ddvpI
+a^2\sum_K\left[
U_{\vp_K\vp_I}+\frac{1}{\Hcal}\left(
\vp_{0I}'U_{\vp_K}+\vp_{0K}'U_{\vp_I}
+\vp_{0K}'\vp_{0I}'\frac{1}{\Hcal}U_0
\right)
\right]\ddvpK \nonumber\\
&+&\frac{2}{\Hcal}\Bigg[\
 \dvpI' \sum_K X_K\dvpK
+\sum_K\vp_{0K}'\dvpK \sum_K a^2 U_{\vp_I\vp_K}\dvpK
\Bigg] \nonumber\\
&+&\left(\frac{1}{\Hcal}\right)^2 \sum_K\vp_{0K}'\dvpK
\Bigg[\
a^2 U_{\vp_I}\sum_K\vp_{0K}'\dvpK
+\vp_{0I}'
\sum_K\left(a^2 U_{\vp_K}+X_K\right)\dvpK
\Bigg]\nonumber\\
&-&2\left(\frac{1}{2\Hcal}\right)^2\frac{\vp_{0I}'}{\Hcal}
\sum_K X_K\dvpK \sum_K \left( X_K\dvpK
+ \vp_{0K}'\dvpK'\right)
+\frac{1}{2\Hcal}\vp_{0I}'\sum_K{\dvpK'}^2
\nonumber\\
&+&a^2\sum_{K,L} \left[
U_{\vp_I\vp_K\vp_L} 
+ \frac{1}{\Hcal}\vp_{0I}' U_{,\vp_K\vp_L}
\right]\dvpK \delta \vp_{1L}
+C\left(\dvpK',\dvpK\right)\nonumber\\
&+&a^2\frac{1}{4}\left\{\frac{dI^2}{d\varphi_{0I}}+\frac{\varphi_{0I}}{\Hcal}I^2\right\}(F^2-\gamma F \tilde F)=0\,
\eea
where $C\left(\dvpK',\dvpK \right)$ contains gradients and inverse
gradients quadratic in the field fluctuations and is defined in the appendix and $X_I=a^2\left(
\frac{1}{\Hcal}U_0\vp_{0I}'+U_{\vp_I}
\right)$. 
The above equation for multiple scalar
fields contains only terms of the field fluctuations $\ddvpI$ and $\dvpI^2$. The detailed derivation appears in Appendix A. 

The result \eqref{flatKG0real},\eqref{flatKG1real},\eqref{flatKG2real} do not assume any slow/fast-roll approximation, and they are general for any source term that appears at second order in the fluctuations. For example coupling fermions in a Yukawa type interaction $y \varphi_I \bar \Psi \Psi$ will also behave similarly, with contributions behaving as $\delta U_2$. The reason is obvious. As long as these additional fields are a negligible amount of the energy density, their sole appearance is in quadratic form. (Otherwise they break Lorentz symmetry.) Thus, these fluctuations are inherently second order and appear as a potential term. 
\be
\mathcal{\tilde V}=f(\vp_{0I})V(\delta^2).
\ee

After this general result, we will simplify our analysis considerably. Our goal is to calculate the scalar spectrum due to sourced fluctuations in a slowly-contracting model of \eqref{eq:action}. For this purpose, it is sufficient to use a \textit{single} scalar field. Furthermore, considering the specific potential 
$U=-U_1e^{-\sqrt{2/p}\varphi}$ causes many potential terms such as $X_I=a^2\left(
\frac{1}{\Hcal}U_0\vp_{0I}'+U_{\vp_I}
\right)=0$ to vanish.  These simplifications lead to:
\bea
\label{eq:singleKG2real}
\ddvp''+2\Hcal\ddvp'-\nabla^2\ddvp
%
%
+\frac{1}{2\Hcal}\vp_{0}'{\dvp'}^2
%
+C\left(\dvp',\dvp\right)
+a^2\frac{1}{4}\left\{\frac{dI^2}{d\varphi_{0}}+\frac{\varphi_{0}}{\Hcal}I^2\right\}(F^2-\gamma F \tilde F)=0.\,\cr
\eea

Note that as in the first order calculation, the potential terms exactly cancel the metric fluctuations. The contribution at second order will only come from first order squared terms such as $\dvp^2$ and the gauge field source term. Using
\be
I(\tau)=(-\tau)^{-n}\equiv a_2^ne^{-n(1-p)/\sqrt{2p}\varphi}
\ee
we see that 
\be
\left\{\frac{dI^2}{d\varphi_{0}}+\frac{\varphi_{0}'}{\Hcal}I^2\right\}=\left(1-n(1-p)\right)\sqrt{\frac{2}{p}} I^2(\varphi_0).
\ee
This will be useful since we will absorb $a^2I^2(\varphi_0)$ into the definitions of the gauge field and get the known Coulomb wave function solutions.
Before carrying out the sourced spectrum calculation, we should address the $\dvp^2$ terms appearing in \eqref{eq:singleKG2real}. These terms appear also in the absence of the gauge field. 

\bea
&&\hspace{-5mm}
\frac{1}{2\Hcal}\vp_{0}'{\dvp'}^2+
C\left(\dvp',\dvp\right)
%
=
\frac{1}{2\Hcal}\vp_{0}'{\dvp'}^2
+\left(\frac{1}{\Hcal}\right)^2
\delta\vp_{1,l}'\nabla^{-2}\left(
\vp_{0}'\dvp'\right)_{,}^{~l}
-2\frac{\vp_{0}'}{\Hcal}\nabla^2\dvp\dvp
\nonumber\\
%
%
&&\hspace{-5mm}
+\frac{\vp_{0}'}{2\Hcal} \delta\vp_{1,l}\delta\vp_{1,}^{~~l}
%
+\left(\frac{\vp_{0}'}{2\Hcal}\right)^2
\frac{\vp_{0}'}{\Hcal}\Bigg[
%
-\delta\vp_{1,l}
\delta\vp_{1,}^{~~~~l}
\Bigg]\nonumber\\
&&\hspace{-5mm}
-\frac{\vp_{0}'}{\Hcal} \nabla^{-2}
\Bigg\{
\left(\delta \vp_{1,l}\nabla^{2}\delta \vp_{1,}^{~~l}
+\nabla^{2}\dvp\nabla^{2}\dvp
+\dvp'\nabla^{2}\dvp'+\delta \vp'_{1,l}\delta \vp_{1,}^{\prime~~l}\right)
-\left(\frac{\vp_0'}{2\Hcal}\right)^2
\Bigg[
+\delta \vp_{1,}^{~~i}\,\delta \vp_{1,j}
\Bigg]_{,i}^{~j}
\Bigg\}\,.\cr
\eea
In a bouncing model, we have a fast-roll rather than slow-roll.
Hence, terms with $\varphi_0'/\Hcal=\sqrt{2/p}\gg1$ dominate, contrary to inflation where they are slow-roll suppressed.
Considering the dominant terms in $\vp_0'/\Hcal$ we expect the last term in the second line and the last term in the third line to be the most dominant. 
Since $\langle \delta \vp_1\delta \vp_1\rangle \sim P_S^v\sim k^2$ we expect these terms to again give a very blue spectrum that is completely irrelevant for large scales.
After neglecting these $\dvp^2$ contributions, we are left with a simple second order differential equation with a source:

\be
\label{eq:KG2source}
\ddvp''+2\Hcal\ddvp'-\nabla^2\ddvp
=-a^2\frac{1}{4}\left[1-n(1-p)\right]\sqrt{\frac{2}{p}} I^2(\varphi_0)(F^2-\gamma F \tilde F).
\ee
\section{Calculation of the Spectrum}
The curvature perturbation is $\zeta=-\frac{H}{\dot{\varphi_0}}\ddvp/M_{pl}=\sqrt{\frac{p}{2}}\ddvp/M_{pl}$.
Performing the transformation of variables to $Q=a\ddvp$ we arrive at the following equation:
\be
Q''-\frac{a''}{a}Q-\nabla^2Q=-\frac{a^3}{4 M_{pl}}\left[1-n(1-p)\right] \sqrt{\frac{p}{2}} I^2(\varphi_0)(F^2-\gamma F \tilde F),
\ee
and in Fourier space:
\be\label{eq:Qequation}
Q_k''+\left(k^2-\frac{a''}{a}\right)Q_k=-\frac{a^3}{4M_{pl}}C (-\tau)^{-2n}(F^2-\gamma F \tilde F)_k,
\ee
where we denoted $\left[1-n(1-p)\right]\sqrt{\frac{2}{p}}=C$, and used $I(\varphi_0)=(-\tau)^{-n}$.
Notice that this factor $C$, that can be traced back to the expression of $I(\tau)$ as a function of $\varphi$, is the only place that breaks the duality of $n\rightarrow -1-n, \xi\rightarrow \xi$. Unlike the tensor spectrum where the duality is exact, here it is broken. 
But because $-\frac{H}{\dot{\varphi_0}}C=(1-n(1-p))$ and $p\ll1$ the effect on $r$ is limited.
 The LHS is the same equation as the tensor perturbation, and therefore the vacuum solution and the retarded Green's function of \eqref{eq:Qequation} are identical to \cite{Ben-Dayan:2016iks} up to the $-\frac{H}{\dot{\varphi_0}}$ normalization:
\begin{align}
G_{k}\left(\tau,\tau'\right) 
=  i\Theta\left(\tau-\tau'\right)\frac{\pi }{4}\sqrt{\tau\tau'}\left[H_{1/2-b}^{(1)}\left(-k\tau\right)H_{1/2-b}^{(2)}\left(-k\tau'\right)-H_{1/2-b}^{(1)}\left(-k\tau'\right)H_{1/2-b}^{(2)}\left(-k\tau\right)\right].
\end{align}

In many bouncing models, the slow-contraction is followed by kinetic domination, with $b=1/2$. At this phase the Green's function vanishes outside the horizon and the production mechanism of sourced fluctuations is shut down. 
As we have seen in \eqref{eq:modeA}, the gauge field fluctuations get amplified at horizon crossing $-k\tau=1$, and we are interested in their amplitude during freeze-out, after horizon exit.  Already from dimensional grounds $E\sim A/\tau$ while $B\sim k A$, hence the $B/E\sim -k \tau \ll1$. 
Therefore the leading behavior of the source term will come from the $\vec E^2$ term:
\be
J\equiv-\frac{a^3}{4M_{pl}}C I^2(\varphi_0)(F^2-\gamma F \tilde F)\simeq -\frac{a^3}{4M_{pl}}C I^2(\varphi_0)(-2)\vec{\hat E}^2=\frac{a^3}{2M_{pl}}CI^2\vec{\hat E}^2
\ee
\bea
\hat E_i^{(\lambda)}(\vec k,\tau)&=&-\frac{1}{a^2}\epsilon_i^{(\lambda)}(\hat k)\partial_{\tau}\hat A_{\lambda}\\
\hat A_i(\tau, k)&=&\sum_{\lambda = \pm}\epsilon_i^{(\lambda)}(\hat k) \frac{ \tilde {A}_{\lambda} (k,\tau)}{I(\tau)} [\hat a_{\lambda}(\vec k)+\hat a^{\dagger}_{\lambda}(-\vec k)]\\
 \tilde {A}_{\lambda} (\tau, \vec q)&\simeq& \sqrt{-\frac{\tau}{2\pi}}e^{\xi\pi}\Gamma(|2n+1|)|2\xi q \tau|^{-|n+1/2|}\\
 J_{\lambda}(\tau, \vec k)&\simeq&\frac{a^3}{2M_{pl}}CI^2\int \frac{d^3p}{(2\pi)^{3/2}}\hat E_i^{(\lambda)}(\vec p,\tau) \hat E_i^{(\lambda)}(\vec k-\vec p,\tau)\cr
& \simeq& \frac{a^3}{2M_{pl}}CI^2\int \frac{d^3p}{(2\pi)^{3/2}}\Bigg[a^{-4}\epsilon_i^{(\lambda)}(\hat p)\epsilon_i^{(\lambda)}\left(\hat{ \vec k-\vec p}\quad \right)\left(\partial_{\tau}\left( \frac{\tilde { A}_{\lambda}}{I}\right)\right)^2\cr
&\times&[\hat a_{\lambda}(\vec p)+\hat a^{\dagger}_{\lambda}(-\vec p)][\hat a_{\lambda}(\vec k-\vec p)+\hat a^{\dagger}_{\lambda}(-(\vec k-\vec p))]\Bigg]\cr
 J_{\lambda}(\tau, \vec k)&=&\frac{C}{2aM_{pl}}\int \frac{d^3p}{(2\pi)^{3/2}}\epsilon_i^{(\lambda)}(\hat p)\epsilon_i^{(\lambda)}\left(\hat{ \vec k-\vec p}\quad \right)[\hat a_{\lambda}(\vec p)+\hat a^{\dagger}_{\lambda}(-\vec p)][\hat a_{\lambda}(\vec k-\vec p)+\hat a^{\dagger}_{\lambda}(-(\vec k-\vec p))]\cr
&\times&\frac{\left(\tilde A'_{\lambda}I -\tilde A_{\lambda}I'\right)^2}{I^2}.\label{eq:scalar_source}
\eea
Thus from equation \eqref{eq:Qequation} the curvature perturbation is given by:
\bea
 \hat \zeta&=&\sqrt{\frac{p}{2}}\frac{\delta \phi_2}{M_p}=\sqrt{\frac{p}{2}}\frac{Q_{\lambda}}{a(\tau)M_p}=\sqrt{\frac{p}{2}}\int^{\tau}d\tau' \frac{G_k(\tau,\tau')}{a(\tau)M_p}J_{\lambda}(k,\tau')\cr
&=&\sqrt{\frac{p}{2}}\frac{C}{2M^2_{pl}}\int^{\tau}d\tau' \frac{G_k(\tau,\tau')}{a(\tau)a(\tau')}\int \frac{d^3p}{(2\pi)^{3/2}}\epsilon_i^{(\lambda)}(\hat p)\epsilon_i^{(\lambda)}\left(\hat{ \vec k-\vec p}\quad \right)[\hat a_{\lambda}(\vec p)+\hat a^{\dagger}_{\lambda}(-\vec p)][\hat a_{\lambda}(\vec k-\vec p)+\hat a^{\dagger}_{\lambda}(-(\vec k-\vec p))]\cr
&\times&\frac{\left(\tilde A'_{\lambda}I -\tilde A_{\lambda}I'\right)^2}{I^2}.
\eea

Now that the curvature perturbation is found, we can calculate the power spectrum. To simplify this, we note that there is a one to one correspondence with the tensor case discussed in \cite{Ben-Dayan:2016iks}.  
The tensor source term is:
	\bea
	J_{\lambda}(\tau,\vec k)&=&\frac{-1}{2M_{pl} a}\int \frac{d^3p}{(2\pi)^{3/2}}\sum_{\lambda'=\pm}\epsilon_i^{(\lambda)*}(\vec k)\epsilon_j^{(\lambda)*}(\vec k)\epsilon_i^{(\lambda')}(\vec p)\epsilon_j^{(\lambda')}(\vec k-\vec p)\cr
	&\times&\left[\hat a_{\lambda}(\vec p)+\hat a^{\dagger}_{\lambda}(-\vec p)\right]\left[\hat a_{\lambda}(\vec k-\vec p)+\hat a^{\dagger}_{\lambda}(-(\vec k-\vec p))\right]\frac{\left(\tilde A'_{\lambda}I -\tilde A_{\lambda}I'\right)^2}{I^2}\label{eq:tensor_source}
	\eea
	where we correct a factor of $1/2$ compared to (7.8) of \cite{Ben-Dayan:2016iks}. 
	Thus, we see that the only difference between the scalar \eqref{eq:scalar_source}
and tensor source \eqref{eq:tensor_source}
is the relevant projection tensor and the factor $C$ in \eqref{eq:scalar_source}. 
	\bea
	\hat h_{\lambda}&=&-\frac{1}{M_{pl}^2}\int^{\tau}d\tau' \frac{G_k(\tau,\tau')}{a(\tau)a(\tau')}\int \frac{d^3p}{(2\pi)^{3/2}} P_{\lambda}(\vec k,\vec p, \vec k-\vec p)[\hat a_{\lambda}(\vec p)+\hat a^{\dagger}_{\lambda}(-\vec p)][\hat a_{\lambda}(\vec k-\vec p)+\hat a^{\dagger}_{\lambda}(-(\vec k-\vec p))]\cr
	&\times&\frac{\left(\tilde A'_{\lambda}I -\tilde A_{\lambda}I'\right)^2}{I^2}
	\eea
	where $P_{\lambda}(\vec k,\vec p, \vec k-\vec p)\equiv\epsilon_i^{(\lambda)*}(\vec k)\epsilon_i^{+}(\vec p)\epsilon_j^{(\lambda)*}(\vec k)\epsilon_j^{+}(\vec k-\vec p)$, because only the $(+)$ polarization is enhanced.
	Thus, both scalar and tensor spectra and their ratio can be written schematically in the following way for $\hat X=\hat \zeta, \hat h$ :
	\bea
	<X_{k}X_{k'}>=\frac{2\pi^{2}}{k^3}\delta(\vec{k}+\vec{k}')({\cal P}^v_X(k)+{\cal P}^s_X(k)).\\
	\mathcal{P}^s_{T,S}=\frac{2{\mathcal N}_m^{T,S\, 2} \mathcal{I}^2_m}{2\pi^2}\frac{e^{4\pi \xi}\xi^{2\alpha}}{M_{pl}^4}k^{6+2\alpha}\times f^{T,S}(q) \label{eq:pts}\\
	r\equiv \frac{\mathcal{P}^s_T}{\mathcal{P}^s_S}=\frac{1}{\left(1-n(1-p)\right)^2}\frac{f^T(q)}{f^S(q)} \label{eq:rexact}
	\eea
	where $\alpha=(-1)^m(2n+1)$, where $m=2$ is used in the $n<-1/2$, and $m=1$ in the $n>-1/2$ case. The numerical factor for the $m=2, n<-1/2$ case is ${\mathcal N}^T_2=\frac{-2 \times4^n (n+1)^2 \Gamma (-2 n-1)^2 }{\pi}
$, and a similar one is obtained by $n\rightarrow -1-n$ for the $m=1, n>-1/2$ case. For the scalar, 
 $\mathcal{N}^S=\sqrt{\frac{p}{2}}C\mathcal{N}^T=(1-n(1-p))\mathcal{N}^T$, and $f^{T},f^{S}$ are the momentum integrals for the tensor
	and scalar terms respectively.  The time dependence integral $\mathcal{I}_m$ is identical to the tensor one.
	\begin{align}
	\label{I2}
	{\cal I}_2&\equiv \int^{\tau}d\tau'\frac{G_k(\tau,\tau')}{a(\tau)a(\tau')}(-\tau')^{2 n}\simeq 
	\frac{1}{k_{end}^{2b}}\left(\frac{\Gamma(1/2-b)\Gamma(1-b+n)}{2^{2b-2n}\Gamma(1/2-n)}k^{-2+2b-2n}-\frac{(-\tau_{end})^{2-2b+2n}}{2(1-b+n)}\right)
	\end{align}
	where $k_{end}=H_{end}/b$. The first term dominates for $-1<n<-1/2$ and the second term for $-2<n<-1$ for $b=p/(1-p)\ll1$ \footnote{A corresponding expression exists for the $-1/2<n<1$ case.}. 
	Hence, the only difference can come from the phase space integration. The tilt of both spectra will be the same.  We are thus led to consider only cases where $n\rightarrow -2$ or equivalently $n\rightarrow 1$, since such $n$ will give us a scale invariant or nearly scale invariant spectrum. Notice that the tilt is still slightly blue, but we assume this can be overcome easily by making the  argument in the exponent of the potential $U=-U_1e^{-\sqrt{2/p}\varphi}$ a slowly varying function of $\varphi$ \cite{Lehners:2010fy}. 
	
	The momentum integrals are
		\bea
	f^{S} & =&\int\frac{d^{3}p}{(2\pi)^{3}}|\epsilon_i\epsilon_i|^{2}(p|\vec{k}-\vec{p}|)^{2n+1}\\
	f^{T} & =&\int\frac{d^{3}p}{(2\pi)^{3}}|P_{\lambda}|^{2}(p|\vec{k}-\vec{p}|)^{2n+1}.
	\eea 
	 where $\epsilon$ is the polarization vector, such that \cite{Sorbo:2011rz,Barnaby:2010vf,Ben-Dayan:2016iks}:
	 \bea
|\epsilon_i^{(\lambda)}(\hat p)\epsilon_i^{(\lambda')}\left(\hat{ \vec k-\vec p}\quad \right)|^2=\frac{1}{4}\left[1-\lambda \lambda' \hat p\left(\hat{ \vec k-\vec p}\quad \right)\right]^2\\
\label{absP}
|P_{\lambda}(\vec k, \vec p, \vec k-\vec p)|^2=\frac{1}{16}\left(1+\lambda \frac{\vec k \cdot \vec p}{kp}\right)^2 \left(1+\lambda \frac{k^2-\vec k \cdot \vec p}{k|\vec k-\vec p|}\right)^2. 
\eea
 The momentum integral can be dealt with using standard dimensional regularization techniques, and does not require a numerical approximation, contrary to previous works.

	 A cumbersome but straightforward calculation yields\footnote{We have already factored out the $k^{5+4n}$ factor, and this factor is a part of the $k^{6+2\alpha}$ in equation \eqref{eq:pts}.}
	\bea
	\label{eq:fs}
	f^{S}&=&\frac{1}{(128 \pi^3 (2 n+1))}\Bigg(\sqrt{\pi} 2^{-4 n} \cos^2(\pi n) \Gamma(-2 n-1/2) \Gamma(2 n+1)\cr
	&-&32 \sin(2\pi n) \left(n (8 n+19)+(n+2) (4 n+3) (4 n+5) \cos(2 \pi n)+12\right) \Gamma(-4 n-6) \Gamma(2 n+2)^2\Bigg) \cr
	\eea
	
	\bea
	\label{eq:ft}
	f^{T}&=&\frac{\Gamma (2 n+1)}{32768 \pi ^3}\Bigg[\frac{976896 \sin (\pi  n) \cos ^3(\pi  n) \Gamma (-4 n-7) \Gamma (2 n+4)}{4
   n+9}\cr
   &-&\Bigg\{\frac{ (n (n (2 n (2 n (2 n (256 n (4 n
   (n+12)+245)+177885)+613759)+1320225)+1728341)+627237))}{\pi ^{-3/2}4^{2 n}(n+1) (n+2) (2 n+1)
   (2 n+3) (2 n+5) \Gamma \left(2 n+\frac{11}{2}\right)}\cr
   &+&\frac{(n (n (2 n (2 n (2 n (32
   n+861)+10831)+58401)+153349)+94821)+21492)}{\pi ^{-3/2}4^{2 n}(n+1) (n+2) (2 n+1)
   (2 n+3) (2 n+5) \Gamma \left(2 n+\frac{11}{2}\right)\cos (2\pi n)}\cr
   &+&\frac{95832}{\pi ^{-3/2}4^{2 n}(n+1) (n+2) (2 n+1)
   (2 n+3) (2 n+5) \Gamma \left(2 n+\frac{11}{2}\right)}\Bigg\}\Bigg].
   \eea
	The important point is the expansion of $r$ around $n=-2$. In this case:
	
	\bea
	r&=&\frac{\left[1+\frac{1}{96} (n+2) \left(2950-1503 \gamma +9 \pi ^2+240 \log (2)-1623 \log (4)-1623 \psi ^{(0)}\left(\frac{3}{2}\right)+120 \psi
		^{(0)}\left(\frac{7}{2}\right)\right)\right]}{\left(1-n(1-p)\right)^{2}}\cr
	&\simeq& \frac{1+1.68(2+n)}{\left(1-n(1-p)\right)^{2}}
	\eea
where $\psi^{(0)}$ is the PolyGamma function.	Note that the leading contribution for the phase space integration is identical, $f^T/f^S=1+\mathcal{O}(n+2)$.
	
	Finally, at the limit of $n\rightarrow -2$ we arrive at:
	\be
	\lim_{n\rightarrow-2}r=\left(\frac{1}{3-2p}\right)^2.\
	\ee
	So for $p\ll1$, we get $r\simeq 1/9$.
	 \footnote{The model discussed in \cite{Ito:2016fqp} corresponds in our analysis to $n=2, \gamma=0$. In such case $f^T/f^S=7/6$ and $r\sim 7/54\sim 0.13$ for $p<<1$ and the tilt is very red with $n_S-1=n_T=4(1-n)=-4$.}
	 The rationale behind the ekpyrotic scenario suggests a phase of stiff matter $w\gg1$ to avoid the anisotropy instability. It seems that once we are in the slow contraction regime of $\frac{2}{3(1+w)}=p \ll 1$, the effect on $r$ is negligible. The lowest possible $r$ is obtained at the $p\rightarrow 0$ limit, and is then strictly $r=1/9$. Any "faster" contraction will result in a larger $r$ still in discord with observations. 

	
	\section{Model Predictions for Present Day Searches}
	The ongoing search of a stochastic GW background using current PTA and LIGO experiments and future LISA mission can also target the dynamics described here. The above analysis showed that $n_S-1=n_T$ and that $rA_S=A_T$, with $r$ given by \eqref{eq:rexact}. Hence, to explain CMB observations we were led to $n_T=0$ that led to $r\simeq1/9$ that is ruled out. However, the calculation did not fix $A_S$. Hence, if the dynamics do not have to account for the observed CMB scalar spectrum, it could still act as a generator of a stochastic background that we may observe with LIGO/LISA/PTA \cite{TheLIGOScientific:2016dpb,Bartolo:2016ami,Kramer:2010tm}. Such a possibility has to fulfill three requirements: \begin{itemize}
	\item The amplitude of both sourced scalar and sourced tensor spectrum are negligible on CMB scales. 
	\item The amplitude of the sourced tensor spectrum is large enough at the relevant LIGO/LISA/PTA scale, such that $\Omega_{GW}$ today is above the future sensitivity curves of these experiments, but lower than the current sensitivity curves.
	\item On all scales the spectrum fulfills the backreaction bound of \eqref{eq:backreaction}.
	\end{itemize}
	As such, the spectra do not have to be nearly scale invariant. For $-2<n<-5/4$ or $1/4<n<1$, we get a tensor tilt of $0<n_T,n_S-1<3$, and its exact functional form is:
	\bea
	\mathcal{P}^s_{T,S}&=&\frac{2{\mathcal N}_m^{T,S\, 2} \mathcal{I}^2_m}{2\pi^2}\frac{e^{4\pi \xi}\xi^{2\alpha}}{M_{pl}^4}k^{6+2\alpha}\times f^{T,S}(q)\\
	\mathcal{P}^s_{T}&=&\frac{11.1 (4-n_T)^2 \Gamma
   \left(3-n_T/2\right)^4}{2^{16-n_T}\pi ^6 n_T}\frac{e^{4\pi \xi}}{b^{4-n_T}\xi^{6-n_T}}\left(\frac{H_{end}}{M_{pl}}\right)^4\left(\frac{k}{H_{end}}\right)^{n_T}\cr
  \label{eq:ptnumerical} &=&\frac{11.1 (4-n_T)^2 \Gamma
   \left(3-n_T/2\right)^4}{2^{16-n_T}\pi ^6 n_T}\frac{e^{4\pi \xi}}{\xi^{6-n_T}}\left(\frac{k_{end}}{M_{pl}}\right)^4\left(\frac{k}{k_{end}}\right)^{n_T}\equiv C(n_T,\xi, k_{end},M_{pl})k^{n_T}\cr
	\eea
	with $f^{T,S}$ as given in \eqref{eq:fs} for the scalar and in \eqref{eq:ft} for the tensor. To simplify the analysis,  in the second line we approximated $f^T \simeq \frac{11.1}{64 \pi^2 (2 + n)}=\frac{11.1}{16 \pi^2 n_T}$ with an accuracy of better than $10\%$ for $-2<n<-1.32$\footnote{The constraints due to the dual branch of $1/4<n<1$ are the same.}. In the third line we substituted $H_{end}=b k_{end}$ that causes $b$ to drop out of the expression of the spectrum. 
 Apart from the requirements mentioned above, there are additional bounds such as BBN constraints and the absence of primordial black holes, but the backreaction constraint is stronger than the others, so fulfilling the backreaction bound is sufficient \cite{Ben-Dayan:2016iks}.
  We shall specify both the spectrum and backreaction bound in terms of the tilt $n_T$, $k_{end}$, that specifies the duration of contraction, and $\xi$ that is in charge of enhancing the spectrum. 
	
	The aforementioned experiments probe the present day fractional energy density stored in stochastic GW.
	The relation between the primordial tensor spectrum to the fractional energy density for $k>k_{eq}$ is given by \cite{Baumann:2007zm}:
	\be \label{eq:omegagw}
	\Omega_{GW}=4.2 \times 10^{-2}\mathcal{P}_T\frac{a_{eq}}{a(\tau_0)}\simeq 4.2 \times 10^{-2}\frac{\mathcal{P}_T}{3400},
	\ee
	since $a(\tau_0)\simeq 3400 a_{eq}$.
	Using
	\be \label{eq:ftok}
k=\frac{2\pi a(\eta_0)f}{c}, 
\ee
the projected forecasts for the different experiments are listed in Table \ref{tableforecast}, \cite{Ben-Dayan:2019gll}.
\begin{table}[h!]
\begin{center}
\caption{Current and forecast detection of the fractional energy density of stochastic GW background by present and  future GW observations. The upper bound is a current constraint and the lower one a future detection threshold.}
\label{tableforecast}
\begin{tabular}{c|c|c}
$Experiment$  & $\Omega^{exp}_{GW}$ & $ k_{exp}\, (Mpc^{-1})$ \\
\hline
$LIGO/aLIGO$ & $1.7\times 10^{-7}>\Omega^{exp}_{GW}> 10^{-9}$ & $3\times 10^{16}-1.3 \times 10^{17}$  \\
$PTA/SKA-PTA$ & $1.3\times 10^{-9}>\Omega^{exp}_{GW}> 1.3 \times 10^{-12}$ & $\sim 1.5 \times10^8$  \\
$LISA$ & $\Omega^{exp}_{GW} >10^{-13}$ & $1.5 \times10^{12}- 1.5 \times 10^{13}$ \\
\end{tabular}
\end{center}
\end{table}
Given that the tensor spectrum is a simple power law, the first two requirements can be phrased in terms of the power spectrum as follows:
\bea
\mathcal{P}_T^s(k_0)<1.3\times 10^{-10},\quad \mathcal{P}_T^s(k_{exp})>0.8 \times 10^{5}\, \Omega_{GW}^{exp}\cr
\Leftrightarrow 0.8 \times 10^{5}\times \Omega_{GW}^{exp} \times \left(\frac{k_0}{k_{exp}}\right)^{n_T}<\mathcal{P}_T^s(k_0)<1.3\times 10^{-10}
\label{ptk0}
\eea
where $k_{exp}$ is the relevant range of wave numbers for each experiment and $\Omega_{GW}^{exp}$ is the minimal detectable fractional density of each experiment.  
The upper bound is from the CMB measurements of $r<0.06$ and $A_S=2.1 \times 10^{-9}$ as reported by PLANCK.
In the analysis below, we shall take $k_0\sim 0.01 Mpc^{-1}, M_{pl}=10^{28}\, Mpc^{-1}$. 
 Substituting the relevant $k_{exp},\Omega_{GW}^{exp}$ such that \eqref{ptk0} is not an empty set, we get a lower bound on $n_T$. For LISA, $k_{LISA}=10^{13}\, Mpc^{-1}$ we get a necessary condition  $n_T>0.12$. The same exercise gives for LIGO $n_T>0.30-0.31$ and for the SKA-PTA $n_T>0.27$.

\begin{figure}
\includegraphics[width=0.5\textwidth]{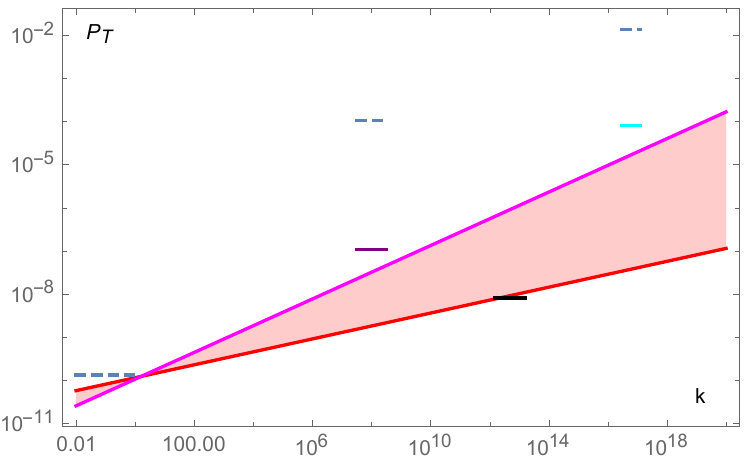}
\includegraphics[width=0.5\textwidth]{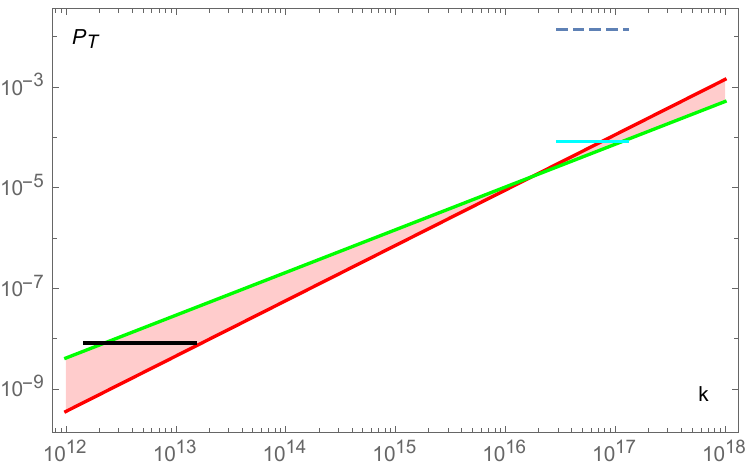}
\begin{center}\includegraphics[width=0.5\textwidth]{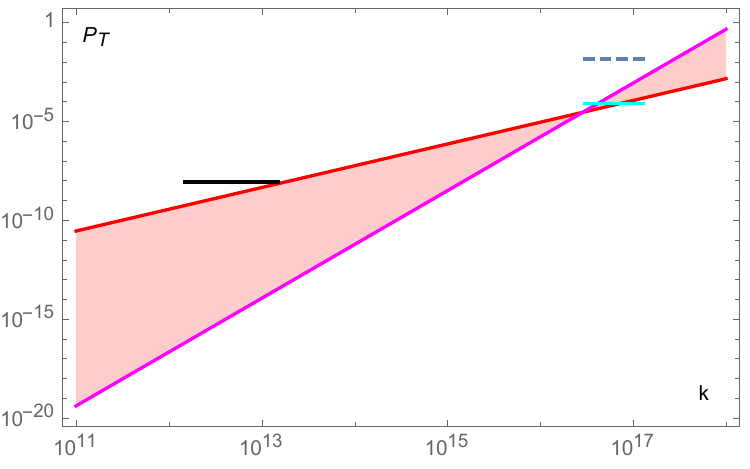} \end{center}
\caption{The GW spectrum  $\mathcal{P}^s_T$ as a function of wavenumber $k$. The shaded region in all plots is allowed by the backreaction bound \eqref{eq:br}. Current observational constraints- CMB, PTA, and LIGO are denoted by dashed lines from left to right. Future sensitivity curves are denoted by lines from left to right, - PTA, (purple), LISA, (black), and LIGO, (cyan). The upper left panel corresponds to LISA only detection (black thick line). The flattest allowed spectrum (red) corresponds to $n_T=0.15,\, \xi=4.55, \,k_{end}=10^{22}\, Mpc^{-1}$ and $b=1/60$, the steepest one (magenta) to  $n_T=0.31, \xi=8.22,\, k_{end}=10^{18}\, Mpc^{-1}$ and $b=1/20$. The upper right panel corresponds to detection by LIGO and LISA. The flattest spectrum (green) corresponds to  $n_T=0.85,\, \xi=8.49$ and the steepest (red) to $n_T=1.1,\xi=8.58$. In this range we always took  $k_{end}=10^{18}\, Mpc^{-1}$ and $b=1/20$. In the lower panel, the LIGO only region $1.1<n_T<2.72$, and $\xi=8.58$ for the flattest spectrum, $\xi=9$ for the steepest one. Again, $k_{end}=10^{18}\, Mpc^{-1}$ and $b=1/20$. }
\label{fig:spectrum}
\end{figure}

Substituting the backreaction bound \eqref{eq:backreaction} for \eqref{eq:ptnumerical} gives:
\bea
\mathcal{P}^s_T= 
\frac{11.1 (4-n_T)^2 \Gamma
   \left(3-n_T/2\right)^4}{2^{16-n_T}\pi ^6 n_T}\frac{e^{4\pi \xi}}{b^{4-n_T}\xi^{6-n_T}}\left(\frac{H_{end}}{M_{pl}}\right)^4\left(\frac{k}{H_{end}}\right)^{n_T}
   \ll  \frac{99.9\times 4 n_T}{ (n_T-4)^2}\xi^{n_T}b^{4}\left(\frac{k}{k_{end}}\right)^{n_T}.\cr
\eea
We see that the larger $b$ is the easier it will be to satisfy the bound.
Substituting the above parameters, the bound is fulfilled if 
\be
\label{eq:br}
\frac{(4-n_T)^4 \Gamma
   \left(3-n_T/2\right)^4}{9\times2^{18-n_T}\pi ^6 n_T^2}\frac{e^{4\pi \xi}}{b^{4}\xi^{6}}\left(\frac{k_{end}}{M_{pl}}\right)^4\ll1.
   \ee
  We see that this can be easily fulfilled for $n_T>0.12$, for a nice range of $\xi,k_{end}$ and $b$. These parameters are interrelated. In our analysis, let us consider $n_T>0.12$ so the spectrum may be detected. For a valid contraction we consider $1/60\leq b\leq 1/20$. $k_{end}$ that determines the duration of the contraction is taken within the range $10^{18}<k_{end}<10^{22}\, Mpc^{-1}$, the lower bound corresponding to at least one more decade beyond the LIGO band, and the upper bound to $60$ e-folds of contraction. Depending on these parameters, $\xi$ is limited to be $\xi<4.5-9$.
  We find three possible regimes of detection - LISA only,  (advanced) LIGO only, and a narrow range of detection by both observations.
  The LISA only regime is characterized by a relatively flat spectrum with $0.15<n_T<0.31$. The mutual regime by $0.85<n_T< 1.1$ and the LIGO only regime by $1.1<n_T<2.72$ \footnote{The $n_T<2.72$ was taken so we could use the approximation for $f^T$. Otherwise, also $2.72<n_T<3$ is valid. Such a blue spectrum with the backreaction bound may only be detected by LIGO.}. The range of spectra is shown in  figure \ref{fig:spectrum}, where we phrase everything in terms of $\mathcal{P}^{s}_T$. The horizontal lines represent current observational constraints (dashed lines) and forecasted sensitivities according to table \ref{tableforecast}, solid lines. Spectra outside the shaded region either violate the backreaction bound, or CMB observations, or are unobservable by any of the experiments assuming their forecasted sensitivity. In all cases $\mathcal{P}_S\simeq\mathcal{P}_T<0.01$ and therefore do not generate primordial black holes, except potentially the last e-fold of the steeper LIGO only case at $k\simeq 10^{18} Mpc^{-1}$. It is clear that the shaded regions show spectra that are not observed by CMB or PTA and that LISA will cover the largest part of the parameter space. 

\section{Discussion}

In this note, we further explored the idea that sourced fluctuations can generate viable CMB spectra, and specifically sizable r in contracting scenarios. For that purpose, we have generalized the second order KG equation to include source terms for any type of cosmology without using slow-roll/fast-roll approximation. If the source term is of second order, it simply modifies the KG equation additively as an additional potential term. Finally, we used an exact solution of the EFE with $U=-U_1e^{\sqrt{2/p} \varphi}$. In such a case the tensor and scalar calculation are almost the same, the sole difference appearing in phase space factors of the momentum integral, and some $\mathcal O (1)$ coefficient. Specifically the spectral tilt is exactly the same, and the amplitude differs by this $\mathcal O (1)$ coefficient. As a result, $r\sim 1/9$ contrary to current bounds. 

Unlike previous works by various authors, it turns out the momentum integral can be carried out exactly using dimensional regularization, which is certainly an interesting technical development. Actually, one may consider using a similar technique to calculate the time integral. If so, it would yield qualitatively different results, as it will remove the power law divergence. However, for the time integral, there is a physical cut-off which is the end of the slow-contraction at $\tau_{end}$, after which kinetic domination starts and the Green's function vanishing outside the horizon, shutting down the sourced fluctuations production mechanism \cite{Ben-Dayan:2016iks}. Hence, this technique is not suitable for the time integral ${\cal I}_m$ that cuts off physically at $\tau=\tau_{end}$. 
The sourced fluctuations bouncing scenario does give a nearly scale invariant, blue, chiral spectrum of gravitational waves, making them a potential target for both CMB and LI experiments. It fails on the quantitative level.
Giving up the idea that the sourced fluctuations bouncing scenario is responsible for CMB results allows us to deviate from near scale invariance and consider potential detection by other experiments such as PTA, LISA and LIGO. We have found that in such case the sourced fluctuations bouncing scenario predicts a GW spectrum that is potentially observable by LISA and LIGO, but not by PTA. 

\subsection*{Acknowledgements}
We thank Merav Hadad and Ruth Shir for useful discussions. The work was partly funded by project RA1800000062.

\appendix
\section{Derivation of the Curvature Perturbation Equation}. 
\numberwithin{equation}{section}
\setcounter{equation}{0}
Here we provide the detailed derivation of \eqref{flatKG2real}. A comma denotes differentiation with respect to the spacetime coordinates.
The metric tensor up to second order for scalar perturbations is
\bea
\label{metric1app}
g_{00}&=&-a^2\left(1+2\phi_1+\phi_2\right) \,, \\
g_{0i}&=&a^2\left(B_1+\frac{1}{2}B_2\right)_{,i}\,, \\
g_{ij}&=&a^2\left[\left(1-2\psi_1-\psi_2\right)\delta_{ij}
+2E_{1,ij}+E_{2,ij}\right]\,.
\eea
and its contravariant form is
\bea
\label{metric2app}
g^{00}&=&-a^{-2}\left[1-2\phi_1-\phi_2+4\phi_1^2-
B_{1,k}B_{1,}^{~k}\right] \,, \\
g^{0i}&=&a^{-2}\left[B_{1,}^{~i}+\frac{1}{2}B_{2,}^{~i}
-2B_{1,k}E_{1,}^{~ki}+2\left(\psi_1-\phi_1\right)B_{1,}^{~i}
\right]\,, \\
g^{ij}&=&a^{-2}\left[\left(1+2\psi_1+\psi_2+4\psi_1^2\right)\delta^{ij}
-\left(
2E_{1,}^{~ij}+E_{2,}^{~ij}-4E_{1,}^{~ik}E_{1,k}^{~~j}
+8\psi_1E_{1,}^{~ij}+B_{1,}^{~i}B_{1,}^{~j}\right)
\right].\cr \label{metric2appij}
\eea
Note that \eqref{metric1app}-\eqref{metric2appij} are without gauge
restrictions, i.e.~no gauge has been specified.
We work in the flat gauge where 
\be
\psi_1=\psi_2=E_1=E_2=0.
\ee

Einstein's field equations (EFE) are given by 
\bea
G_{\mu \nu}&=&T_{\mu \nu}\\
%
\label{eq:emtexplicit}
T_{\mu \nu}&=&\sum_K\left(\varphi_{K,\mu}\varphi_{K,\nu}-\frac{1}{2}g_{\mu \nu}g^{\alpha \beta}\varphi_{K,\alpha}\varphi_{K,\beta}\right)-g_{\mu \nu}\left[U(\varphi_K)+\frac{I^2(\varphi_K)}{4}(F^2-\gamma F \tilde F)\right].
\eea
The scalar fields is expanded up to second order according to
\be
\vp_K(\tau,\vec x)=\vp_{0K}(\tau)+\delta \vp_{1K}(\tau, \vec x)+\frac{1}{2}\delta \varphi_{2K}(\tau,\vec x).
\ee
The potential is expanded accordingly,
\be
U(\varphi_I)=U_0+\delta U_1+\frac{1}{2}\delta U_2, \quad \delta U_1=\sum_K U_{\varphi_K} \delta \vp_{1K}, \quad \delta U_2=\sum_{KL}U_{\vp_L \vp_K}{\delta \vp}_{1K}{\delta \vp}_{1L}+\sum_KU_{\varphi_K}\delta \vp_{2K}.
\ee

The zeroth order and first order are unchanged with respect to \cite{Malik:2006ir}. The deviation occurs at second order, where the gauge fields appear.	
\subsection{Zeroth Order}\label{0th}
The zeroth order equations  $0-0$ component $\delta_{ij}$ and KG respectively are simply:
\bea
3\mathcal{H}^2=\frac{1}{2}\sum_I\varphi_{I0}^{'2}+a^2U_0\cr
\mathcal{H}^2-2\frac{a''}{a}=\frac{1}{2}\sum_I\varphi_{I0}^{'2}-a^2U_0\cr
\varphi_{I0}''+2\mathcal{H} \varphi_{I0}'+a^2U_{\varphi_I}=0.
\eea

\subsection{First Order}\label{1st}
Starting from the KG equation at first order,
\be
\delta \vp_{1I}''+2\Hcal\delta \vp_{1I}'+2 a^2U_{\varphi_I}\phi_1-\nabla^2\delta \vp_{1I}
-\varphi_{0I}'\nabla^2B_1-\varphi_{0I}'\phi_1'+a^2\sum_K
U_{\vp_K\vp_I}
\delta \vp_{1K}=0\,.
\ee
Using the background equations, the $0-0$
component at first order will be:
\be
\label{00Ein1}
2a^2U_0{\phi_1}+\sum_K\vp_{0K}'{\dvp}_K'+a^2{\delta U_1}
+2\Hcal \nabla^2B_1
=0\,,
\ee
the $0-i$ part gives
\be
\label{0iEin1}
\Hcal\phi_1-\frac{1}{2}\sum_K\vp_{0K}'{\dvp}_K=0\,.
\ee
{}From the $i-j$ component of the Einstein equation we get the trace free part
\be
\label{offtrace1}
B_1'+2\Hcal B_1+\phi_1=0\,.
\ee
Using \eqref{offtrace1} and the zeroth order equations, we get the
first order trace:
\be
\label{trace1}
\Hcal\phi_1'+\frac{1}{2}\left[ 
a^2\delta U_1+2a^2U_0{\phi_1}-\sum_K\vp_{0K}'{{\dvp}_K}'\right]
=0\,.
\ee
%

Using the EFE, we get an equation which has no metric fluctuations in it, but just field fluctuations and background quantities:
\be
\label{KG1_flat}
\delta \vp_{1I}''+2\Hcal\delta \vp_{1I}'-\nabla^2\delta \vp_{1I}
+a^2\sum_K\left\{
U_{\vp_K\vp_I}
+\frac{1}{\Hcal}\left(
\vp_{0I}'U_{\vp_K}+\vp_{0K}'U_{\vp_I}
+\vp_{0K}'\vp_{0I}'\frac{1}{\Hcal}U_0
\right)
\right\}\delta \vp_{1K}=0\,
\ee
and for single field:
\be
\delta \vp_{1}''+2\Hcal\delta \vp_{1}'-\nabla^2\delta \vp_{1}
+a^2\left\{
U_{\vp \vp}
+\frac{1}{\Hcal}\left(
2\vp_{0}'U_{\vp}
+\frac{\vp_{0}'^2}{\Hcal}U_0
\right)
\right\}\delta \vp_{1}=0\,.
\ee

\subsection{Second Order}
To make the crucial algebraic manipulations transparent, we keep the terms $\delta U_2$ as such in the KG equation:
  \bea
\ddvpI''&+&2\Hcal\ddvpI'-\nabla^2\ddvpI+a^2 \frac{\partial \delta U_2}{\partial \vp_{0I}} +a^2\frac{1}{4}\frac{\partial I(\vp_{0I})^2}{\partial\varphi_{0I}}\left(F^2-\gamma F \tilde F\right)+2a^2 U_{,\vp_I}\phi_2
-\vp_{0I}'\left(\nabla^2 B_2+\phi_2'\right)\nonumber\\
&+&4\vp_{0I}' B_{1,k}\phi_{1,}^{~k}
+2\left(2\Hcal\vp_{0I}'+a^2 U_{\vp_I}\right) B_{1,k}B_{1,}^{~k}
+4\phi_1\left(a^2\sum_K U_{\vp_I\vp_K}\delta \vp_{1K}-\nabla^2\delta \vp_{1I}\right)
+4\vp_{0I}'\phi_1\phi_1'\nonumber\\
&-&2\delta \vp_{1I}'\left(\nabla^2 B_1+\phi_1'\right)-4\delta \vp'_{1I,k}B_{1,}^{~k}
=0.
\eea
Substituting $\delta U_2$ gives:
 \bea
\label{KG2flatsingle}
\ddvpI''&+&2\Hcal\ddvpI'-\nabla^2\ddvpI+a^2 \sum_K U_{\vp_I\vp_K}\ddvpK
+a^2 \sum_{K,L} U_{\vp_I\vp_K\vp_L} \delta \vp_{1K}\delta \vp_{1L} +2a^2 U_{,\vp_I}\phi_2
-\vp_{0I}'\left(\nabla^2 B_2+\phi_2'\right)\nonumber\\
&+&4\vp_{0I}' B_{1,k}\phi_{1,}^{~k}
+2\left(2\Hcal\vp_{0I}'+a^2 U_{,\vp_I}\right) B_{1,k}B_{1,}^{~k}
+4\phi_1\left(a^2\sum_K U_{\vp_I\vp_K}\delta \vp_{1K}-\nabla^2\delta \vp_{1I}\right)
+4\vp_{0I}'\phi_1\phi_1'\nonumber\\
&-&2\delta \vp_{1I}'\left(\nabla^2 B_1+\phi_1'\right)-4\delta \vp'_{1I,k}B_{1,}^{~k}
=-\frac{1}{4}\frac{\partial I(\vp_{0I})^2}{\partial \varphi_{0I}}\left(F^2-\gamma F \tilde F\right)\,.
\eea

Now we need to use the field equations to substitute the metric perturbations.  Considering \eqref{eq:emtexplicit}, we see that the gauge field term is already second order in the perturbations. Hence, the metric in front of it will be zeroth order. Therefore, the only change in the energy-momentum tensor will come from diagonal terms. The change in the diagonal terms will behave as an additional 'potential' term, i.e. wherever there is a $\frac{1}{2}\delta U_2$ term it has to be replaced with $\frac{1}{2}\delta U_2+\frac{I^2(\varphi)}{4}(F^2-\gamma F \tilde F)$.
The $0-0$ component at second order gives:

\bea
\label{Ein00_2}
 a^2 U_0\left(\phi_2+B_{1,k}B_{1,}^{~k}\right)
&+&\Hcal\nabla^2B_2 
+\frac{1}{2}\left[B_{1,kl}B_{1,}^{~kl}
-\left(\nabla^2 B_1\right)^2\right]
-2\Hcal \phi_{1,k}B_{1,}^{~k}+a^2\frac{I^2(\varphi_{0I})}{4}(F^2-\gamma F \tilde F)\nonumber\\
&+&\frac{1}{2}\sum_K\left[
\vp_{0K}'\ddvpK' +a^2\delta U_2+4 a^2\delta U_1\phi_1
+{\dvpK'}^2+\delta \vp_{K,k}\delta \vp_{K,}^{~~k}
\right]=0.
\eea
For future reference we see that the equation looks like:
\be
\label{eq:B2}
\nabla^2B_2=\frac{1}{\Hcal}\left[\cdots-a^2\frac{1}{2}\delta U_2-a^2\frac{I^2(\varphi_{0I})}{4}(F^2-\gamma F \tilde F)\right].
\ee
The $0-i$ Einstein equation is the same as in the absence of gauge fields:
\be
\label{0i_2}
\Hcal\phi_{2,i}-4\Hcal\phi_{1}\phi_{1,i}+2\Hcal B_{1,ki}B_{1,}^{~k}
+B_{1,ki}\phi_{1,}^{~k}-\nabla^2 B_1\phi_{1,i}
-\frac{1}{2} \sum_K\left[\vp_{0K}'\delta \vp_{2K,i}+2\dvpK'\delta \vp_{1K,i}\right]=0\,.
\ee
Using first order $0-i$ and taking the trace gives:
\bea
\label{0i_2version2}
&&\Hcal\left(\phi_{2}-2\phi_{1}^2+ B_{1,k}B_{1,}^{~k}\right)
-\frac{1}{2} \sum_K\vp_{0K}'\ddvpK\nonumber\\
&&\qquad
+\nabla^{-2}\left(\phi_{1,kl}B_{1,}^{~~kl}
-\nabla^2 B_1\nabla^2\phi_1\right)
- \sum_K \nabla^{-2}\left(
\dvpK'\nabla^{2}\dvpK+\delta\vp_{1K,l}'\delta\vp_{1K,}^{~~~~l}
\right)
=0,\cr
\eea
where we introduce the inverse Laplacian,
$\nabla^{-2}(\nabla^{2})X=X$.
Let us rewrite the equation as an expression for $\phi_2$:
\bea
\label{eq:phi2}
\phi_{2}&=&-\left(-2\phi_{1}^2+ B_{1,k}B_{1,}^{~k}\right)
+\frac{1}{\Hcal}\frac{1}{2} \sum_K\vp_{0K}'\ddvpK\cr
%
&-&\frac{1}{\Hcal}\nabla^{-2}\left(\phi_{1,kl}B_{1,}^{~~kl}
+\nabla^2 B_1\nabla^2\phi_1\right)
+\frac{1}{\Hcal}\sum_K \nabla^{-2}\left(
\dvpK'\nabla^{2}\dvpK+\delta\vp_{1K,l}'\delta\vp_{1K,}^{~~~~l}
\right).
\eea
%
%
The $i-j$ Einstein equation is given by
\bea
\label{ij2}
&&\Bigg\{
2 a^2 U_0\left(\phi_2-4\phi_1^2+B_{1,k}B_{1,}^{~k}\right)
+2\Hcal\phi_2'-8\Hcal\phi_1\phi_1'-2\phi_{1,k}\phi_{1,}^{~k}
+4\Hcal B_{1,k}'B_{1,}^{~~k}\nonumber\\
&&\qquad
+\nabla^2\left(B_2'+2\Hcal B_2+\phi_2\right)-2\phi_1'\nabla^2 B_1
+B_{1,kl}B_{1,}^{~kl}-\left(\nabla^2 B_1\right)^2
+4\left(a^2 \delta U_1+2 a^2 U_0\phi_1\right)\phi_1\nonumber\\
&&\qquad
+\left[ 
\sum_K\left(
\delta \vp_{1K,l}\delta \vp_{1K,}^{~~~~l}-\vp_{0K}'\ddvpK'-{\dvpK'}^2\right)
+a^2\delta U_2+a^2\frac{I^2(\varphi_{0I})}{2}(F^2-\gamma F \tilde F)\right]\Bigg\}\delta^i_{~j}\\
&&\qquad
-\left(B_2'+2\Hcal B_2+\phi_2\right)^{~i}_{,~j}
+2\phi_{1,}^{~~i}\,\phi_{1,j}+2B_{1,~j}^{~~i}\left(\phi_1'+\nabla^2 B_1\right)
-2B_{1,~k}^{~~i}B_{1,~j}^{~~k}
-2\sum_K\delta \vp_{1K,}^{~i}\delta \vp_{1K,j}=0\,.\nonumber
\eea
After additional manipulations outlined in  \cite{Malik:2006ir}, we finally reach a trace equation for the spatial EFE:
\bea
\label{spatialtrace2}
&&3 a^2 U_0\left(\phi_2-4\phi_1^2+B_{1,k}B_{1,}^{~k}\right)
-2\phi_{1,k}\phi_{1,}^{~k}
-2\phi_1'\nabla^2B_1
-6\Hcal B_{1,}^{~k} \left(2\Hcal B_{1,k}+\phi_{1,k}\right)\nonumber\\
&&+3\Hcal\phi_2'-12\Hcal\phi_1\phi_1'
+\nabla^2\left(B_2'+2\Hcal B_2+\phi_2\right)
+\frac{1}{2}\left[B_{1,kl}B_{1,}^{~kl}-\left(\nabla^2 B_1\right)^2\right]
+6\phi_1\left(a^2 \delta U_1+2 a^2 U_0\phi_1\right)
\nonumber\\
&&+\frac{1}{2}\sum_K\left( 3a^2\delta U_2+3a^2\frac{I^2(\varphi)}{2}(F^2-\gamma F \tilde F)-3\vp_{0K}'\ddvpK'-3{\dvpK'}^2
+\delta \vp_{1K,l}\delta \vp_{1K,}^{~~~~l}\right)
=0\,.
\eea

We now substitute all the metric perturbations in the KG eqation \eqref{KG2flatsingle}. First, substituting only the second order potentials, we use \eqref{eq:B2},\eqref{eq:phi2} leading to:
\bea
\label{flatKG2_intermediate}
\ddvpI''&+&2\Hcal \ddvpI'-\nabla^2\ddvpI+a^2\sum_K U_{\vp_I\vp_K}\ddvpK
+a^2\sum_{K,L} U_{\vp_I\vp_K\vp_L}\dvpK\delta \vp_{1L}+a^2\frac{1}{4}\frac{dI(\vp_0)^2}{d\varphi_0}\left(F^2-\gamma F \tilde F\right)
\cr&+&4\phi_1\left(\sum_K a^2 U_{\vp_I\vp_K}\dvpK-\nabla^2\dvpI\right)
-4\delta\vp_{1I,k}'B_{1,}^{~~k}+4a^2 U_{,\vp_I}\phi_1^2
+\frac{2}{\Hcal}\dvpI'\Xkdvk
\nonumber\\
&+&\frac{1}{\Hcal}\vp_{0I}'\left\{
2\phi_1\left(a^2\delta U_1+\Xkdvk\right)
+\frac{1}{2}\sum_K\left({\dvpK'}^2
+\delta\vp_{1K,l}\delta\vp_{1K,}^{~~~~l}\right)
\right\}\nonumber\\
&+&2\frac{X_I}{\Hcal}\,\nabla^{-2}
\left\{
\nabla^2 B_1\nabla^2\phi_1
-\phi_{1,kl}B_{1,}^{~~kl}
+\sum_K\Big(\dvpK'\nabla^2\dvpK
+\delta\vp_{1K,l}'\delta\vp_{1K,}^{~~~~l}\Big)
\right\}\nonumber\\
&+&\frac{\vp_{0I}'}{\Hcal}\left\{
B_{1,kl}B_{1,}^{~~kl}-\phi_{1,k}\phi_{1,}^{~~k}
-\left(\nabla^2B_1\right)^2
-\left(\frac{1/2}{\Hcal}\right)^2\left[
\left(\Xkdvk\right)^2-\left(\sum_K\vp_{0K}'\dvpK'\right)^2\right]\right\}
\nonumber\\
&-&\frac{\vp_{0I}'}{\Hcal}
\; \nabla^{-2}\Bigg\{
\sum_K\Big(
\delta\vp_{1K,l}\nabla^{2}\delta\vp_{1K,}^{~~~~l}
+\nabla^{2}\dvpK1\nabla^{2}\dvpK
+\dvpK'\nabla^{2}\dvpK'
+\delta\vp_{1K,l}'\delta\vp_{1K,}^{\prime~~~l}\Big)
\nonumber\\
&&\qquad\qquad\qquad
+\Big(
\frac{1}{\Hcal}B_{1,~j}^{~~i}\, \Xkdvk-\phi_{1,}^{~~i}\phi_{1,j}
\Big)_{,i}^{~j}
\Bigg\}
\cr
&+&\frac{1}{\Hcal}\left\{\vp_{0I}' a^2\left[\delta U_2+\frac{I^2(\varphi_{0I})}{4}(F^2-\gamma F \tilde F)\right]
+X_I \sum_K \vp_{0K}'\ddvpK
\right\}
=0\,
\eea
where we have defined:
$X_I\equiv a^2\left(
\frac{1}{\Hcal}U_0\vp_{0I}'+U_{\vp_I}
\right)$.
Now, we again use the first and zeroth order equations to reach a final expression:

\bea
\label{flatKG2realapp}
\ddvpI''&+&2\Hcal\ddvpI'-\nabla^2\ddvpI
+a^2\sum_K\left[
U_{\vp_K\vp_I}+\frac{1}{\Hcal}\left(
\vp_{0I}'U_{\vp_K}+\vp_{0K}'U_{\vp_I}
+\vp_{0K}'\vp_{0I}'\frac{1}{\Hcal}U_0
\right)
\right]\ddvpK \nonumber\\
&+&\frac{2}{\Hcal}\Bigg[\
 \dvpI' \sum_K X_K\dvpK
+\sum_K\vp_{0K}'\dvpK \sum_K a^2 U_{\vp_I\vp_K}\dvpK
\Bigg] \nonumber\\
&+&\left(\frac{1}{\Hcal}\right)^2 \sum_K\vp_{0K}'\dvpK
\Bigg[\
a^2 U_{\vp_I}\sum_K\vp_{0K}'\dvpK
+\vp_{0I}'
\sum_K\left(a^2 U_{\vp_K}+X_K\right)\dvpK
\Bigg]\nonumber\\
&-&2\left(\frac{1}{2\Hcal}\right)^2\frac{\vp_{0I}'}{\Hcal}
\sum_K X_K\dvpK \sum_K \left( X_K\dvpK
+ \vp_{0K}'\dvpK'\right)
+\frac{1}{2\Hcal}\vp_{0I}'\sum_K{\dvpK'}^2
\nonumber\\
&+&a^2\sum_{K,L} \left[
U_{\vp_I\vp_K\vp_L} 
+ \frac{1}{\Hcal}\vp_{0I}' U_{,\vp_K\vp_L}
\right]\dvpK \delta \vp_{1L}
+C\left(\dvpK',\dvpK\right)\nonumber\\
&+&a^2\frac{1}{4}\left\{\frac{dI^2}{d\varphi_{0I}}+\frac{\varphi'_{0I}}{\Hcal}I^2\right\}(F^2-\gamma F \tilde F)=0\,
\eea
where $C\left(\dvpK',\dvpK \right)$ contains gradients and inverse
gradients quadratic in the field fluctuations and is defined as
%
%
\bea
\label{Fdvk1}
&&\hspace{-5mm}
C\left(\dvpK',\dvpK\right)
%
=
\left(\frac{1}{\Hcal}\right)^2
\delta\vp_{1I,l}'\nabla^{-2}\sum_K\left(
X_K\dvpK+\vp_{0K}'\dvpK'\right)_{,}^{~l}
-\frac{2}{\Hcal}\nabla^2\dvpI\sum_K\vp_{0K}'\dvpK
\nonumber\\
&&\hspace{-5mm}
+2\frac{X_I}{\Hcal}\left(\frac{1}{2\Hcal}\right)^2 
\nabla^{-2}
\Bigg[
\sum_K\vp_{0K}'\delta\vp_{1K,lm}
\nabla^{-2}\sum_K\left(X_K\dvpK+\vp_{0K}'\dvpK'\right)_{,}^{~lm}
\nonumber\\
&&-\sum_K\left(X_K\dvpK+\vp_{0K}'\dvpK'\right)
\nabla^{2}\sum_K\vp_{0K}'\dvpK
\Bigg]
\nonumber\\
&&\hspace{-5mm}
+\frac{1}{2\Hcal}
\Bigg[\vp_{0I}'\sum_K \delta\vp_{1K,l}\delta\vp_{1K,}^{~~~~l}
+4X_I\nabla^{-2}\sum_K\left(
\dvpK'\nabla^2\dvpK+\delta \vp'_{1K,l}\delta \vp_{1K,}^{~~~~l}\right)
\Bigg]
\nonumber\\
&&\hspace{-5mm}
+\left(\frac{1}{2\Hcal}\right)^2
\frac{\vp_{0I}'}{\Hcal}\Bigg[
\nabla^{-2}\sum_K\left( X_K\dvpK+\vp_{0K}'\dvpK'\right)_{,lm}
\nabla^{-2}\sum_K\left(X_K\dvpK+\vp_{0K}'\dvpK'\right)_{,}^{~lm}
\nonumber\\
&-&\sum_K\vp_{0K}'\delta\vp_{1K,l}
\sum_K\vp_{0K}'\delta\vp_{1K,}^{~~~~l}
\Bigg]\nonumber\\
&&\hspace{-5mm}
-\frac{\vp_{0I}'}{\Hcal} \nabla^{-2}
\Bigg\{
\sum_K\left(\delta \vp_{1K,l}\nabla^{2}\delta \vp_{1K,}^{~~~~l}
+\nabla^{2}\dvpK\nabla^{2}\dvpK
+\dvpK'\nabla^{2}\dvpK'+\delta \vp'_{1K,l}\delta \vp_{1K,}^{\prime~~~~l}\right)
\nonumber\\
&&\qquad
-\left(\frac{1}{2\Hcal}\right)^2
\Bigg[
2\nabla^{-2}\sum_K\left(X_K\dvpK+\vp_{0K}'\dvpK'\right)_{,~j}^{~i}
\sum_K X_K\dvpK
+\sum_K\vp_{0K}'\delta \vp_{1K,}^{~~~i}\sum_K\vp_{0K}'\delta \vp_{1K,j}
\Bigg]_{,i}^{~j}
\Bigg\}\,.\cr
\eea

\section{Example of the momentum integral calculation}
In this appendix we give an example of how we evaluated the momentum integral exactly. The integral to be solved is
\begin{equation}
f^{S}=\int\frac{d^{3}p}{\left(2\pi\right)^{3}}\left|\epsilon_{i}\epsilon_{i}\right|^{2}\left(p\left|\vec{k}-\vec{p}\right|\right)^{2n+1}\label{eq:fs2}
\end{equation}
where 
\begin{equation}
\left|\epsilon\left(\hat{p}\right)\epsilon\left(\vec{k}-\vec{p}\right)\right|^{2}=\frac{1}{4}\left(1-\hat{p}\left(\vec{k}-\vec{p}\right)\right)^{2}.\label{eq:epsilon term}
\end{equation}
First we rewrite equation \eqref{eq:epsilon term} in a more tractable form.

\begin{equation}
1-2\frac{\vec{p}\cdot\vec{k}-p^{2}}{p(\vec{k}-\vec{p})}+\frac{\left(\vec{p}\cdot\vec{k}-p^{2}\right)^2}{p^{2}(\vec{k}-\vec{p})^{2}}
\end{equation}
 where $\vec{p}\cdot\vec{k}=(p^{2}+k^{2}-(\vec{k}-\vec{p})^{2}/2$.
So equation \eqref{eq:fs2} is
\begin{equation}
\frac{1}{4}\int\frac{d^{3}p}{\left(2\pi\right)^{3}}\left|\vec{k}-\vec{p}\right|^{2n+1}p^{2n+1}\left(1-2\frac{\vec{p}\cdot\vec{k}-p^{2}}{p(\vec{k}-\vec{p})}+\frac{\left(\vec{p}\cdot\vec{k}-p^{2}\right)}{p^{2}(\vec{k}-\vec{p})^{2}}\right).\label{eq:FINAL INTEGRAND}
\end{equation}
The integrand expands into nine terms as follows. We write $K\equiv \left|\vec{k}-\vec{p}\right|$ and obtain
\begin{align}
 & -k^{2}K^{2n}p^{2n}+K^{2+2n}p^{2n}+K^{2n}p^{2+2n}+\frac{1}{4}K^{2n-1}p^{2n+3}-\frac{1}{2}k^{2}K^{2n-1}p^{2n+1}+\nonumber \\
+ & \frac{3}{2}K^{2n+1}p^{2n+1}+\frac{1}{4}k^{4}K^{2n-1}p^{2n-1}-\frac{1}{2}k^{2}K^{2n+1}p^{2n-1}+\frac{1}{4}K^{3+2n}p^{2n-1}.\label{eq:big integrand}
\end{align}
In order to solve the integral we use the Schwinger parameter representation:
\begin{equation}
\frac{1}{\left(k^{2}-m^{2}\right)^{n}}=\left(-i\right)^{n}\frac{1}{\Gamma\left(n\right)}\intop_{0}^{\infty}d\alpha\alpha^{n-1}e^{i\alpha\left(k^{2}-m^{2}\right)}.\label{eq:schwinger}
\end{equation}
As an example we calculate the first term of the momentum integral,$-k^{2}\left|\vec{k}-\vec{p}\right|^{2n}p^{2n}$
which is a product of two expressions like \eqref{eq:schwinger}.
We write $-n=n_{1}=n_{2}$ where we keep $n_{1},n_{2}$ distinct to
clarify how the other eight terms are worked out: 
\begin{equation}
\int d^{3}p\frac{1}{p^{2n_{1}}\left(p-k\right)^{2n_{2}}}=\frac{\left(-i\right)^{n_{1}+n_{2}}}{\Gamma\left[n_{1}\right]\Gamma\left[n_{2}\right]}\int d^{3}p\intop_{0}^{\infty}d\alpha_{1}\alpha_{1}^{n_{1}-1}e^{-i\alpha_{1}p^{2}}\intop_{0}^{\infty}d\alpha_{2}\alpha_{2}^{n_{2}-1}e^{i\alpha_{2}\left(p-k\right)^{2}}.
\end{equation}
The integral over p can be converted into a Gaussian integral for
p, after some algebra:
\begin{align}
\int d^{3}p\frac{1}{p^{2n_{1}}\left(p-k\right)^{2n_{2}}} & =\frac{\left(-i\right)^{n_{1}+n_{2}}}{\Gamma\left[n_{1}\right]\Gamma\left[n_{2}\right]}\intop_{0}^{\infty}d\alpha_{1}d\alpha_{2}\alpha_{1}^{n_{1}-1}\alpha_{2}^{n_{2}-1}\nonumber \\
\times & \intop\frac{d^{3}p}{\left(2\pi\right)^{3}}e^{i\left(\alpha_{1}+\alpha_{2}\right)\left(p+\frac{\alpha_{2}}{\alpha_{1}+\alpha_{2}}k\right)^{2}}e^{-i\frac{\alpha_{2}^{2}}{\alpha_{1}+\alpha_{2}}k^{2}}e^{i\alpha_{2}k^{2}}
\end{align}
and solving this gives
\begin{equation}
\frac{\left(-i\right)^{n_{1}+n_{2}-3/2}\pi^{3/2}}{\Gamma\left[n_{1}\right]\Gamma\left[n_{2}\right]}\intop_{0}^{\infty}d\alpha_{1}d\alpha_{2}\alpha_{1}^{n_{1}-1}\alpha_{2}^{n_{2}-1}\left(\alpha_{1}+\alpha_{2}\right)^{-3/2}e^{i\frac{\alpha_{1}\alpha_{2}}{\alpha_{1}+\alpha_{2}}k^{2}}.
\end{equation}
The next step is to substitute $\alpha_{1}=\beta t,$ $\alpha_{2}=\beta(1-t)$
and $\alpha_{1}+\alpha_{2}=\beta$, so that we have
\begin{align}
 & \frac{\left(-i\right)^{n_{1}+n_{2}-3/2}\pi^{3/2}}{\Gamma\left[n_{1}\right]\Gamma\left[n_{2}\right]}\intop_{0}^{1}dtt^{n_{1}-1}\left(1-t\right)^{n_{2}-1}\intop_{0}^{\infty}d\beta\beta^{n_{1}+n_{2}-3/2-1}e^{i\beta t(1-t)k^{2}}.
\end{align}
 Substituting $x=\beta t(1-t)k^{2}$, the integral over $\beta$ is
seen to be a Gamma function, for which the solution is known, and
we obtain
\begin{align}
 & \frac{\pi^{3/2}\Gamma\left[n_{1}+n_{2}-\frac{3}{2}\right]}{\Gamma\left[n_{1}\right]\Gamma\left[n_{2}\right]}(k^{2})^{3/2-n_{1}-n_{2}}\intop_{0}^{1}dtt^{3/2-n_{2}-1}\left(1-t\right)^{3/2-n_{1}-1}.
\end{align}
The integral over t is a Beta function, so finally we obtain

\begin{align}
\int d^{3}p\frac{1}{p^{2n_{1}}\left(p-k\right)^{2n_{2}}} & =\frac{\pi^{3/2}\Gamma\left[n_{1}+n_{2}-\frac{3}{2}\right]\Gamma\left[\frac{3}{2}-n_{1}\right]\Gamma\left[\frac{3}{2}-n_{2}\right]}{\Gamma\left[n_{1}\right]\Gamma\left[n_{2}\right]\Gamma\left[3-n_{1}-n_{2}\right]}(k^{2})^{3/2-n_{1}-n_{2}}.\\
\nonumber 
\end{align}
We now insert $n_{1}=n_{2}=-n$ and put back the prefactors $-k^{2}/\left(2\pi\right)^{3}$
, so this term becomes
\begin{equation}
-k^{5+4n}\frac{\Gamma\left[-2n-\frac{3}{2}\right]\Gamma\left[\frac{3}{2}+n\right]^{2}}{\pi^{3/2}\Gamma\left[-n\right]^{2}\Gamma\left[3+2n\right]}.
\end{equation}
 The other eight terms have an identical structure to the first, and
are solved in exactly the same way, so we finally obtain
\begin{align}
f^{S} & =\frac{k^{5+4n}}{(128\pi^{3}(2n+1))}\Bigg(\sqrt{\pi}2^{-4n}\cos^{2}(\pi n)\Gamma(-2n-1/2)\Gamma(2n+1)%
\nonumber \\
 & -32\sin(2\pi n)\left(n(8n+19)+(n+2)(4n+3)(4n+5)\cos(2\pi n)+12\right)\Gamma(-4n-6)\Gamma(2n+2)^{2}\Bigg).
\end{align}
The same method is used to obtain the result for $f^T$ written in \eqref{eq:ft}.

\end{document}